\documentclass[aps,prd,nofootinbib,preprintnumbers,twocolumn,showpacs,floatfix]{revtex4-1}

\usepackage[T1]{fontenc}
\usepackage{amsmath}
\usepackage{amssymb}
\usepackage{graphicx}
\usepackage{color}
\usepackage{hyperref}

\newcommand{\bitem}{\begin{itemize}}
\newcommand{\eitem}{\end{itemize}}
\newcommand{\bwt}{\begin{widetext}}
\newcommand{\ewt}{\end{widetext}}
\newcommand{\be}{\begin{equation}}
\newcommand{\ee}{\end{equation}}

\newcommand\g{\gamma}
\newcommand{\bdm}{\begin{displaymath}}
\newcommand{\edm}{\end{displaymath}}
\newcommand{\bea}{\begin{eqnarray}}
\newcommand{\eea}{\end{eqnarray}}

\newcommand{\nn}{\nonumber}

\renewcommand\H{\mathcal H}
\renewcommand\P{\mathcal P}
\newcommand\C{\mathcal C}
\newcommand\e{\mathrm e}
\newcommand\sand[3]{\big\langle\overline{#1#2}\left|#3\right|{#1#2}\big\rangle}
\def\eq#1{{Eq.~(\ref{#1})}}
\def\eqs#1#2{{Eqs.~(\ref{#1})--(\ref{#2})}}
\def\fig#1{{Fig.~\ref{#1}}}
\def\figs#1#2{{Figs.~\ref{#1}--\ref{#2}}}
\def\Table#1{{Table~\ref{#1}}}

\def\abs#1{\left| #1\right|}
\def\mod#1{\abs{#1}}
\def\Im{\mbox{Im}\,}
\def\Re{\mbox{Re}\,}

\def\dtheta#1#2{{\theta_{#1}-\theta_{#2}}}

\newcommand\TeV{\text{TeV}}

\def\Eq#1{{Eq.~\ref{#1}}}
\def\Eqs#1#2{{Eqs.~\ref{#1}--\ref{#2}}}

\begin{document}
\jot = 1.2ex         

\preprint{}
\title{Present and Future $K$ and $B$ Meson Mixing Constraints on TeV Scale Left-Right Symmetry}
\author{Stefano Bertolini}\email{stefano.bertolini@sissa.it}
\affiliation{INFN, Sezione di Trieste, SISSA,
via Bonomea 265, 34136 Trieste, Italy}
\author{Alessio Maiezza}\email{alessio.maiezza@ific.uv.es}
\affiliation{IFIC, Universitat de Val\`encia-CSIC, Apt. Correus 22085, E-46071 Val\`encia, Spain}
\author{Fabrizio Nesti}\email{fabrizio.nesti@irb.hr}
\affiliation{Ru\dj er Bo\v skovi\'c Institute, Bijeni\v cka cesta 54, 10000, Zagreb, Croatia}
\affiliation{Gran Sasso Science Institute, viale Crispi 7, I-67100 L'Aquila, Italy}
\begin{abstract}
\noindent
We revisit the $\Delta F=2$ transitions in the $K$ and $B_{d,s}$ neutral meson systems in the
context of the minimal Left-Right symmetric model. We take into account, in addition to up-to-date
phenomenological data, the contributions related to the renormalization of the flavor-changing
neutral Higgs tree-level amplitude. These contributions were neglected in recent discussions, albeit
formally needed in order to obtain a gauge independent result.  Their impact on the minimal LR model
is crucial and twofold. First, the effects are relevant in $B$ meson oscillations, for both CP
conserving and CP violating observables, so that for the first time these imply constraints on the LR
scenario which compete with those of the $K$ sector (plagued by long-distance uncertainties).
Second, they sizably contribute to the indirect kaon CP violation parameter $\varepsilon$.  We
discuss the bounds from $B$ and $K$ mesons in both cases of LR symmetry: generalized parity ($\P$)
and charge conjugation ($\C$).  In the case of $\P$, the interplay between the CP-violation
parameters $\varepsilon$ and $\varepsilon'$ leads us to rule out the regime of very hierarchical
bidoublet vacuum expectation values $v_2/v_1<m_b/m_t\simeq 0.02$.  In general, by minimizing the
scalar field contribution up to the limit of the perturbative regime and by definite values of the
relevant CP phases in the charged right-handed currents, we find that a right-handed gauge boson
$W_R$ as light as $3\,$TeV is allowed at the 95\%\,CL. This is well within the reach of direct
detection at the next LHC run.  If not discovered, within a decade the upgraded LHCb and Super B
factories may reach an indirect sensitivity to a Left-Right scale of 8\,TeV.
\end{abstract}

\maketitle

\section{Introduction}

\noindent
The Left-Right (LR) symmetric extension of the Standard Model (SM)~\cite{Pati:1974yy,
  Mohapatra:1974hk, Mohapatra:1974gc, Senjanovic:1975rk, Senjanovic:1978ev}, provide a natural setup
for understanding the origin of parity violation as well as the smallness of neutrino masses via the
see-saw mechanism~\cite{Minkowski:1977sc, Gell-Mann:1980vs, Yanagida:1979as, Glashow:1979nm,
  Mohapatra:1979ia}, which intrinsically connects the two energy scales.  Such a framework has been
revived in the recent years for its potential collider implications when parity restoration in the
LHC energy reach is considered. Intriguing is the possibility that neutrinoless-double-beta-decay
($0\nu2\beta$) may be dominated by the $W_R$ gauge boson exchange~\cite{Feinberg:1978,
  Mohapatra:1980yp, Tello:2010am} and therefore lead to a signal even when the improving
cosmological limit on the light neutrino masses~\cite{Hannestad:2010yi, Archidiacono:2013cha} may
prevent them to be responsible for it.  This has a direct counterpart in the Keung-Senjanovic
process at colliders where the very same lepton number violation can appear as same-sign
leptons~\cite{Keung:1983uu}, constituting a clean signal of the right-handed (RH) gauge boson $W_R$,
with very low background.  The LR a setup has further the capability of addressing also the dark
matter issue in a predictive scenario~\cite{Nemevsek:2012cd}.  All this fertile framework triggered
a number of authors to investigate both direct and indirect signatures of a TeV scale RH gauge
interaction as well as constraints from flavor changing processes~\cite{Zhang:2007da,
  Maiezza:2010ic, Guadagnoli:2010sd, Blanke:2011ry, Nemevsek:2011aa, Nemevsek:2012iq, Dev:2013oxa,
  Dev:2013vba, Huang:2013kma, Roitgrund:2014zka, Bertolini:2012pu, Barry:2012ga,
  AguilarSaavedra:2012fu, Das:2012ii, Han:2012vk, Krasnikov:2013ifa, Barry:2013xxa, Kou:2013gna,
  Bertolini:2013noa}.  Flavor and CP violating loop processes provide a sensitive and powerful
testground for any extension of the SM. For the minimal LR model, in Ref.~\cite{Maiezza:2010ic} an
absolute lower bound for the LR scale of $\sim 2.5$\,TeV was obtained, in full reach of LHC direct
searches. As in earlier studies, such result came essentially from the constraint on new physics
contributions to $\Delta{M_K}$.

\begin{figure*}[t]
\centerline{\includegraphics[width=14cm]{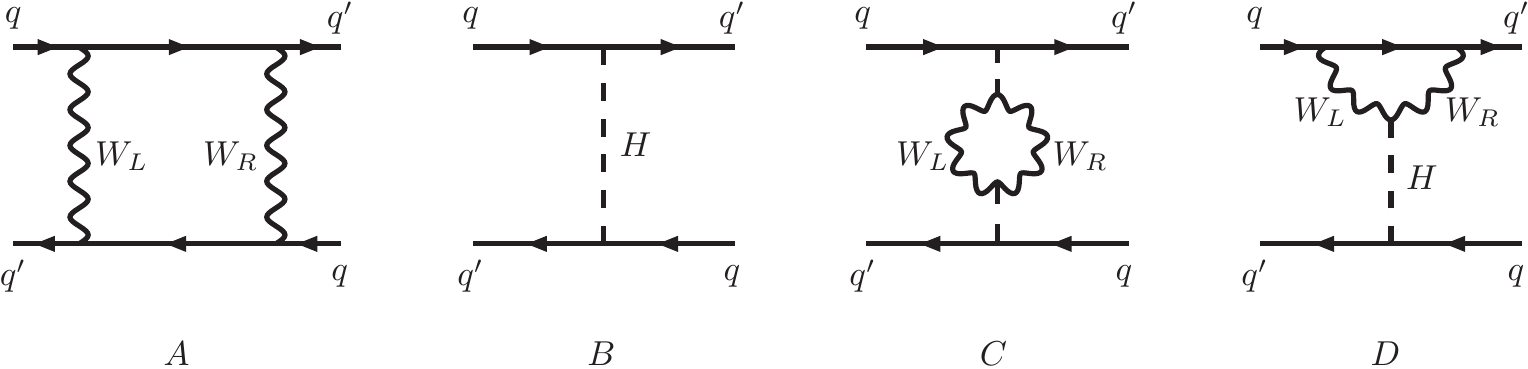}}%
\caption{The four classes of leading LR diagrams contributing to the neutral meson mixings. A,B,C,D
  identify the box diagrams, the $H$-mediated tree-level amplitude, the one-loop self-energy and
  vertex renormalizations, respectively. The diagrams drawn are just representatives of each class
  and all allowed contractions among external and internal states are understood. In classes C and D
  all diagrams which do not contain the $HW_{L}W_{R}$ vertex (with the longitudinal components of
  the gauge bosons) are subleading for large $M_H$, analogously to $H$ exchange in the box A.
  \label{fig:Diagrams}\vspace*{-1ex}}
\end{figure*}

In the present paper we focus again on $\Delta F=2$ transitions of $K$ and $B$ mesons. Besides
updating experimental data, we improve on previous analyses in two crucial respects. First, together
with the LR box and the tree-level flavor-changing (FC) Higgs amplitudes, we include the leading
one-loop LR renormalizations of the latter, which were neglected in recent discussions but are
needed in order to obtain a gauge independent result~\cite{Chang:1984hr, Basecq:1985cr,Hou:1985ur,
  Ecker:1985vv}. These additional contributions add constructively and play a relevant role in the
total amplitude. Secondly, we improve on the assessment of the QCD renormalization factors.  In
particular, the coefficient of the amplitudes with top and charm quarks exchanged in the loop was
underestimated in Ref.~\cite{Maiezza:2010ic}.

This has a relevant implication for CP violation in $K$ mixing where the charm-top contribution
plays a crucial role in the LR model.  The destructive interference between the $cc$ and $ct$
amplitudes, achieved by a given configuration of the relevant LR phases, is in fact more efficient
than estimated in the past. As a consequence, given present data and the the related uncertainties,
$\Delta B=2$ mixing and related CP violation play now a leading role in constraining low scale LR
symmetry.  We expect this feature to become even more prominent in the future with more data coming
from LHCb and B-factories, while only a substantial theoretical improvement on the calculation of
the long-distance contributions to the $K_L$-$K_S$ mass difference may make this observable prevail
over $B$ data.

In minimal LR models a discrete symmetry is often assumed that relates the couplings in the left and
right sectors, the only two realistic implementations being generalized parity ($\mathcal{P}$) and
generalized charge-conjugation ($\mathcal{C}$)~\cite{Chang:1983fu,Chang:1984uy,Maiezza:2010ic}.  The
latter arises naturally in a grand unified SO(10) embedding as a generator of the
algebra~\cite{Senjanovic:2011zz,Arbelaez:2013nga}.  We analyze the impact of meson oscillations in
both cases of low scale symmetry restoration.

A first outcome of our analysis is that, in the case of $\P$-parity, after considering the improved
renormalization and the new contributions to CP violation in the $K$ sector ($\varepsilon$ and
$\varepsilon'$) we can strongly rule out the regime of hierarchical VEVs of the bidoublet,
$v_2/v_1<m_b/m_t\simeq 0.02$.  For not so hierarchical VEVs, the predictivity of the model and the
strict correlation among the LR phases requires a fully numerical analysis with constraints from $K$
and $B$ oscillations.

The numerical analysis leads to a reassessment of the absolute lower bounds on the LR scales. We
find the FC Higgs to be bounded by $B$ oscillations to be always above 20\,TeV.  Thus, as it is well
known~\cite{Senjanovic:1979cta}, in order to obtain TeV scale LR symmetry the $W_R$ gauge boson has
to be substantially lighter than the second Higgs doublet, posing the concern of perturbativity of
the scalar coupling to the longitudinal gauge boson
components~\cite{Guadagnoli:2010sd,Blanke:2011ry,Dev:2013vba}. Our result is that, keeping $M_H$ at
the limit of the perturbative regime, a fit of the present $B_d$ and $B_s$ mixing data allows for
$W_R$ as light as $2.9\,(3.2)\,\TeV$ at the 95\%\,CL in the $\C\,(\P)$ case.

The possibility of such a low scales of LR symmetry, favourable to LHC direct detection, is achieved
in both frameworks for quite specific patterns of the model CP phases, which we discuss in detail.

We discuss finally the foreseen improvements following from the constraints on $\Delta B=2$
observables in the upgraded stages of LHCb and Super-B factories, and conclude that they will raise
the sensitivity to the LR scale beyond 6 TeV, thus setting a challenging benchmark for the direct
search (the latter being sensitively dependent on the decay channels and the mass scale of the
right-handed neutrinos, for a discussion see~\cite{Maiezza:2010ic, Nemevsek:2011hz}).

\section{LR model and Meson Oscillations}\label{sec:Hamiltonian}

In the minimal LR model additional $\Delta F=2$ transitions are mediated by the
right-handed gauge boson $W_R$ and the neutral flavor changing Higgs
(FCH) $H$.  We review the model and the relevant lagrangian interactions in
Appendix~\ref{appendix:LR}. Here we just recall that the $W_R$ charged-current
interactions are characterized by a flavor mixing matrix $V_R$ , that is the analogue of the standard
Cabibbo-Kobayashi-Maskawa (CKM) matrix $V_L$.

While the $W_R$ gauge bosons appear in loop diagrams, the FCH mediates $\Delta F=2$ transitions at
the tree level.  Both $W_R$ and $H$ exhibit the same fermion mixing structure, proportional to
$V_{L}^* V_R$ (see App.~\ref{appendix:LR}).  The phenomenological analysis of
Ref.~\cite{Maiezza:2010ic} shows that the mixing angles in $V_R$ are very close to $V_L$, thus
making the model very predictive. To the detail, the precise form $V_R$ depends on the discrete LR
symmetry realization, denoted by $\mathcal{P}$ parity and $\mathcal{C}$ conjugation,
\be
\label{eq:VR}
\P:\ \  V_R\simeq K_uV_LK_d\,,\quad
\C:\ \  V_R = K_uV_L^*K_d\,.
\ee
$K_{u,d}$ are diagonal matrices of phases, namely
$K_u=\mathop{\mathrm{diag}}\{\e^{i\theta_u},\e^{i\theta_c},\e^{i\theta_t}\}$,
$K_{d}=\mathop{\mathrm{diag}}\{\e^{i\theta_d},\e^{i\theta_s},\e^{i\theta_b}\}$.
For a detailed discussion see~\cite{Maiezza:2010ic}. It is enough to recall that in the case of
$\C$, the additional CP phases are independent parameters, while in the case of $\P$ they are
related since the theory has just one free parametric phase beyond the CKM one. In the latter case
an analytic solution was provided in Ref.~\cite{Zhang:2007da} which holds however in a specific
limit of the model lagrangian parameters (see also App.~\ref{appendix:LR}).

In the case of $\C$ the freedom of the CP phases plays a crucial role in evading the stringent
constraints from flavor physics. Recent detailed discussions include Refs.~\cite{Maiezza:2010ic} and
\cite{Bertolini:2012pu}.

\subsection{Effective $\Delta F=2$ LR Hamiltonian }

In Fig.~\ref{fig:Diagrams} all the relevant classes of LR Feynman diagrams for meson oscillations
are shown.  The relevant lagrangian interactions are summarized in the appendix \ref{appendix:LR}.
The four contributions are identified as box, tree-level flavor changing Higgs (FCH) amplitude and
the related self-energy and vertex LR renormalization .

The diagrams drawn are representative of each class, all allowed contractions being understood.  The
Feynman amplitude $A$ in Fig. \ref{fig:Diagrams} is not gauge independent but the sum of the A, B
and C amplitudes does \cite{Chang:1984hr,Basecq:1985cr,Ecker:1985vv}.  Apparently the box and the
C,D diagrams depend on different parameters (namely the $H$ mass).  On the other hand, the Higgs
coupling to the charged would-be-Goldstone-bosons is proportional to $M_{H}^2/M_{W_R}$.  This leads
to a compensation of the Higgs propagator and to contributions independent on the Higgs mass.  The
consequences are twofold. On the one hand the gauge dependence of the box diagram is thereby
canceled; on the other hand even for a Higgs heavier than $W_R$ (such a setup is enforced by the
presence of the tree level FCH amplitude and it is relevant to our discussion) contributions from C
and D arise that are competing in size with the box amplitude. The presence of these contributions
affect sizably as we shall se B-physics observables.

Let us just mention that the corresponding loop diagrams with $W_R$ replacing $W_L$ are suppressed by $M_{W_L}^2/M_{W_R}^2$ compared to those in Fig. \ref{fig:Diagrams}.

The calculation of the the diagrams $A,B,C,D$ gives at low energy the following
effective Hamiltonians~\cite{Basecq:1985cr, Ecker:1985vv}:
\begin{widetext}
\begin{align}
\label{Ha}&\mathcal{H}_ {A}= \frac{2 G^2_F  \beta}{\pi^2} \sum_{i,j} m_i m_j \lambda_{i}^{LR}\lambda_{j}^{RL} \eta_{ij}^A \ F_{A}(x_i,x_j,\beta) \ {O}_S \\
\label{Hb}&\mathcal{H}_ {B}= -\frac{2 \sqrt{2} G_F}{M_H^2} \sum_{i,j} m_i m_j \lambda_{i}^{LR} \lambda_{j}^{RL}   \eta_{ij}^B \ {O}_S \\
\label{Hc}&\mathcal{H}_ {C}= -\frac{G^2_F \beta}{2\pi^2 M_H^2} \sum_{i,j} m_i m_j \lambda_{i}^{LR} \lambda_{j}^{RL}  \eta_{ij}^C \ F_{C}(M_{W_{L}},M_{W_{R}},M_H) \ {O}_S \\
\label{Hd}&\mathcal{H}_ {D}= -\frac{4 G^2_F \beta}{\pi^2 M_H^2} \sum_{i,j} m_i m_j \lambda_{i}^{LR} \lambda_{j}^{RL}  \eta_{ij}^D \ F_{D}(m_{i},m_{j},M_{W_{L}},M_{W_{R}},M_H) \ {O}_S
\end{align}
\end{widetext}
where $\beta=M_{W_L}^2/M_{W_R}^2$, $\lambda_{i}^{LR}=V^{L*}_{id'}V^R_{id}$, $x_i=m_i^2/M_{W_L}^2$
and $i,j=c,t$. The dimension six operator ${O}_S$ is identified with $\bar{s} L d\, \bar{s} R d$,
$\bar{b} L d\, \bar{b} R d$ and $\bar{b} L s\, \bar{b} R s$ for the $K$, $B_{d}$ and $B_{s}$ meson
mixings respectively, with $R,L=(1\pm \gamma_5)/2$.  The coefficients $\eta_{ij}$ encode the effect
of the QCD renormalization down to the relevant hadronic scale (where the matrix elements of ${O}_S$
are evaluated). Finally, the loop functions $F_{A,C,D}$ are defined in
App.~\ref{sec:loopfunctions}.

A few comments are in order.  A complete operator basis for the Hamiltonians in \eqs{Ha}{Hd}
includes $O_V=\bar{d'} \gamma_\mu L d\, \bar{d'} \gamma^\mu R d$. At the leading order (LO) in the
QCD resummation the anomalous dimension matrix diagonalizes
on two multiplicative renormalized operators, proportional respectively to $O_S$ and to
$O_{\widetilde V}=O_V+\tfrac{2}{3}\,O_S$~\cite{Ecker:1985vv}. We verified that the QCD induced amplitude related to
$O_{\tilde V}$ remains in all cases negligibly small compared to that of $O_S$ (for $K$ mixing the
hadronic matrix element of $O_S$ is chirally enhanced as well~\cite{Frere:1991db}).  We neglect the
QCD renormalization above the top mass, since it amounts to a fraction always
below $10\%$ of the whole effect. This amounts to effectively matching the amplitudes at the weak
scale and it allows us to write the QCD corrected Hamiltonian in the simple form of
\eqs{Ha}{Hd}.

The two vertex ($D$) contractions with the loop enclosing the upper and bottom quark line
respectively, lead to an exchange of the L/R chirality in ${O}_S$. An analogous effect has the
interchange of $W_L$ and $W_R$ in the box ($A$) and vertex ($D$) diagrams, while it does not affect the self-energy diagram.
This amounts just to a multiplicity factor for the $\Delta F=2$ transitions we are considering since the operator ${O}_S$ is symmetric for
$L\leftrightarrow R$ when external momenta are neglected.

Finally, a convenient subtraction must be applied to the divergent amplitudes $C$ and $D$ such that $M_H$
identifies with the one-loop pole mass \cite{Basecq:1985cr}.

The numerical relevance of the diagrams $C$ and $D$ compared to $A$ depends on whether Kaon or $B$
meson mixings are considered. As a matter of fact, when the charm quark dominates the $\Delta F=2$
amplitude (for instance when computing the CP conserving $\Delta M_K$) since the box amplitude is
enhanced by a large log (namely $\log(m_c^2/M_{W_L}^2)$) with respect to the amplitudes $C$ and
$D$. For such a component of the amplitude we find that the contribution to $\Delta M_K$ of
$\mathcal{H}_{C+D}$ is confined to be below 20\% of $\mathcal{H}_{A}$.

This is no longer true for CP violating observables in the Kaon system, as $\varepsilon$, or $B$
meson observables, where the top quark exchange leads the loop amplitudes and all diagrams compete.

\subsection{QCD renormalization}\label{sec:renorm}

The effective Hamiltonians are written in \eqs{Ha}{Hd} to hold at the scales relevant to the
considered mesonic transitions, namely $m_b$ for $B$ and the GeV scale for kaons. The effective
four-quark operators receive important QCD renormalization in their evolution from the fundamental
scales.  Not all needed renormalization factors are available at the next-to-leading (NLO) order for
the LR Hamiltonians  here discussed. On the other hand, the QCD renormalization from the left-right scales
down to the weak scale (e.g $m_t$) is readily estimated at LO from Ref.~\cite{Ecker:1985vv} to be a small fraction of the overall
QCD correction (always below 10\%). By neglecting it and matching effectively the LR amplitudes at the weak
scale, the NLO $\eta$-factors for the top quarks mediated diagrams (whose integration leads at the
weak scale to $O_S$) are obtained from Ref.~\cite{Buras:2001ra}.  This is all what is needed
for the discussion of $B^0$-$\overline{B^0}$ mixings, where top exchange dominates the box end
the vertex diagrams, A and D respectively.

Having integrated out the heavy $H$ scalar and the LR gauge boson states, the NLO QCD renormalization
of the tree level FCH (B) and the self-energy diagram C is straightforward and can be obtained from
Ref.~\cite{Buras:2001ra}.  As a matter of fact, the running up-quark masses present in the
flavor-changing $Hdd'$ couplings (see App.~\ref{sec:loopfunctions}) absorb the QCD renormalization
of $O_S$ down to the decoupling quark scale, leaving the residual QCD renormalization of the
effective operator down to the hadronic scale (the LO anomalous dimension of $O_S$ is minus twice
that of a mass).

The case of the box diagrams with one or two intermediate charm quarks can be handled according to
the procedure described in~\cite{Ecker:1985vv, Vysotsky:1979tu}, and partly by using the results of
Ref.~\cite{Buras:2001ra}. We verified that, when both calculations can be compared (e.g. for the t-t
amplitudes), implementing the NLO running coupling in the LO approach of~\cite{Ecker:1985vv}
approximates well (within 20\%) the NLO results given in~\cite{Buras:2001ra}.  In such a case we use
the NLO values derived from Ref.~\cite{Buras:2001ra}.

The QCD renormalization of the vertex diagram (D) with internal charm can be evaluated analogously.
The absence of large logs in the Wilson coefficient (see App.~\ref{sec:loopfunctions}) leads, in the
LO approach of Ref.~\cite{Ecker:1985vv}, to a QCD correction identical to that of the B and C
diagrams.

The numerical values of the $\eta_{QCD}$ coefficients thereby obtained are reported in
\Table{table:eta}.  Since the LO anomalous dimension of ${O}_S$ equals up to the sign that of
$m_i^2$, a large part of the QCD renormalization is absorbed by the running quark masses in the $m_i
m_j$ pre-factor. This justifies the size pattern of the QCD renormalization factors in the table.
The errors due to the uncertainties in the input parameters (strong coupling and mass thresholds)
are as well reduced by the same mechanism amounting to a maximum of 10\% in $\eta_{Ktt}$ and of 5\%
in $\eta_{Btt}$~\cite{Buras:2001ra}.  These uncertainties are included in the conservative ranges we
shall consider for the LR contributions to the relevant observables.  It is worth noting that well
within the uncertainty of the~\cite{Ecker:1985vv} LO calculation ($\alpha_s^{NLO}$ improved) the box
$\eta_{cc,ct}$ coefficients are identical to the corresponding coefficients of the $H$ self-energy and
vertex amplitudes.

\begin{table}
\vspace*{-1.5ex}%
$$\begin{array}{lcccccc}
\hline
                      & \eta_{K cc} &  \eta_{K ct,tc} & \eta_{K tt} & \eta_{B cc} &  \eta_{B ct,tc} & \eta_{B tt}    \\[0.5ex]
\hline
         {\rm A}  &  1.15   &  2.23  &  5.63  &  0.52  & 1.01  & 2.25  \\
  {\rm B,C,D}  &  1.26  &  2.66   &  5.63  &  0.50  & 1.10  & 2.25  \\
\hline
\end{array}$$
\caption{QCD renormalization factors for kaon and $B$ mixing as described in the text.  They are computed at $\mu=1$ GeV and $m_b$ for $K$ and $B$ respectively for central values of the parameters. The observables here discussed are mainly sensitive to $\eta_{K cc,ct}$ and $\eta_{B tt}$.}
\label{table:eta}
\end{table}

\subsection{Hadronic matrix elements}

The hadronic matrix elements of the operators ${O}_S$ can be readily evaluated by factorization via
the vacuum saturation approximation (VSA). One obtains
\bea
 \sand{K}{^0}{\bar{s} L d\, \bar{s} R d}
&=&
\frac12 f_K^2 m_K {\mathcal B}^K_4\left[\frac{m_K^2}{(m_s+m_d)^2} + \frac{1}{6}\right] \nn \\
 \sand{B}{_d^0}{\bar{b} L d\, \bar{b} R d}
&=&
\frac12 f_{B_d}^2 m_{B_d} {\mathcal B}^{B_d}_4\left[\frac{m_{B_d}^2}{(m_b+m_d)^2} + \frac{1}{6}\right] \nn \\
\sand{B}{_s^0}{\bar{b} L s\, \bar{b} R s}
&=&
\frac12 f_{B_s}^2 m_{B_s} {\mathcal B}^{B_s}_4\left[\frac{m_{B_s}^2}{(m_b+m_s)^2} + \frac{1}{6}\right]\nn\\
\label{ME}
\eea
where $f_{K,B_d,B_s}$ and $m_{K,B_d,B_s}$ are the decay constants and the masses of the mesons $K$
and $B_{d,s}$ respectively. The bag factors ${\mathcal B}^M_4$ parametrize the deviation from the
naive VSA.  The first unquenched lattice determinations have appeared in
2012~\cite{Boyle:2012qb,Bertone:2012cu}.  A more recent lattice calculation using staggered fermions
has found discrepant results~\cite{Bae:2013tca}. In particular, a value of ${\mathcal B}^M_4$ about
50\% larger. The origin of this discrepancy is being currently investigated~\cite{Bae:2013tca}. In
\Table{table:inputs} we report the values we use in our analysis~\cite{Aoki:2013ldr}.  The term
$1/6$ in \eq{ME} is numerically subleading and it is often neglected (in the $B$-$\bar B$ matrix
elements as well).  This is taken accordingly into account when using the lattice bag factors in our
numerical analysis.

The quark masses appearing in the matrix elements are scale dependent and they are evaluated at the
relevant hadronic scales.  It is worth noting that by considering the scale dependent VSA matrix elements
 the LR bag factors turn out with very good approximation scale independent. This is related to the $m^{-2}$ anomalous dimension
 of ${O}_S$.

\begin{table}
\vspace*{-1.5ex}%
$$
\begin{array}{ll}
\hline
\text{Parameters}\ \ \ \    &  \text{Input values} \\
\hline
m_t (m_t)  &  164(1)\ \text{GeV} \\
m_b(m_b)   &  4.18(3)\ \text{GeV} \\
m_c(m_c)  &  1.28(3)\ \text{GeV} \\
m_s(2\ \text{GeV})   & 0.095(5)\ \text{GeV} \\
m_s(1\ \text{GeV})  &  0.127(7)\ \text{GeV} \\
{\mathcal B}^{K}_{4}(2\ \text{GeV})    &   0.78(3)    \\
{\mathcal B}^{B_{d}}_{4}(m_b)    &  1.15(3)  \\
{\mathcal B}^{B_{s}}_{4}(m_b)     &  1.16(2)      \\
\hline
\end{array}
$$
\vspace*{-2ex}
\caption{Running quark masses and relevant bag parameters used in the computation.
  The numerical values are given at the NLO in the $\overline{\rm MS}(\rm NDR)$ scheme.
  Errors in the last figures are reported in the round brackets.\vspace*{-1ex}}%
\label{table:inputs}
\end{table}

\section{Constraints from $K$ and $B_{d}$, $B_{s}$ Oscillations}

\subsection{Parametrization of LR amplitudes}

\noindent
For both $K$ and $B_{d,s}$ oscillations, it is useful to discuss the allowed NP constraints in terms
of ratios of the additional contributions to the correspond SM quantities or experimental
data~\cite{Silva:1996ih,Grossman:1997dd,Deshpande:1996yt,Lenz:2006hd,Charles:2005ckm}. We introduce
the parameters
\bea
h^K_{m}&\equiv&\frac{2\ \Re\sand{K}{^0}{\H_{LR}}}{\left(\Delta M_K\right)_{exp}}\, , \label{hKm} \\
h^K_{\varepsilon}&\equiv&\frac{\Im\sand{K}{^0}{\H_{LR}}}{\Im\sand{K}{^0}{\H_{LL}}}\, , \label{hKe}\\
h^B_q&\equiv&\frac{\sand{B}{_q^0}{\H_{LR}}}{\sand{B}{_q^0}{\H_{LL}}}\,, \label{hBq}
\eea
where $\H_{LR}=\H_{A}+\H_{B}+\H_{C}+\H_{D}$ and $q=d,s$.  The SM hamiltonian $\H_{LL}$ is reported
in Appendix \ref{sec:loopfunctions}.  The parameters $h^B_{q}$ are complex, while
$h^K_{m,\varepsilon}$ are real.

\begin{figure}[t]
\centerline{%
\includegraphics[width=0.92\columnwidth]{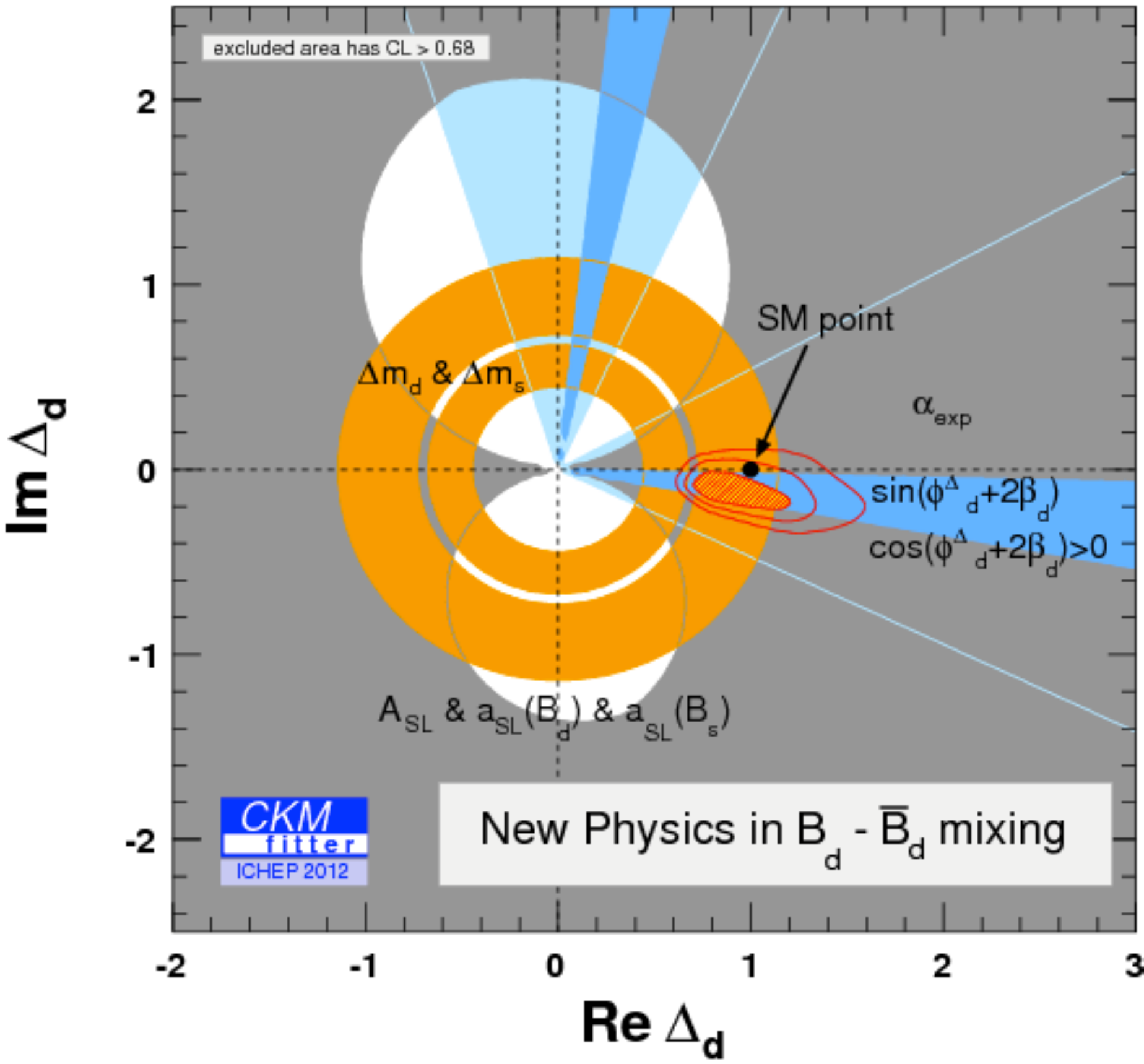}}

\centerline{%
\includegraphics[width=0.92\columnwidth]{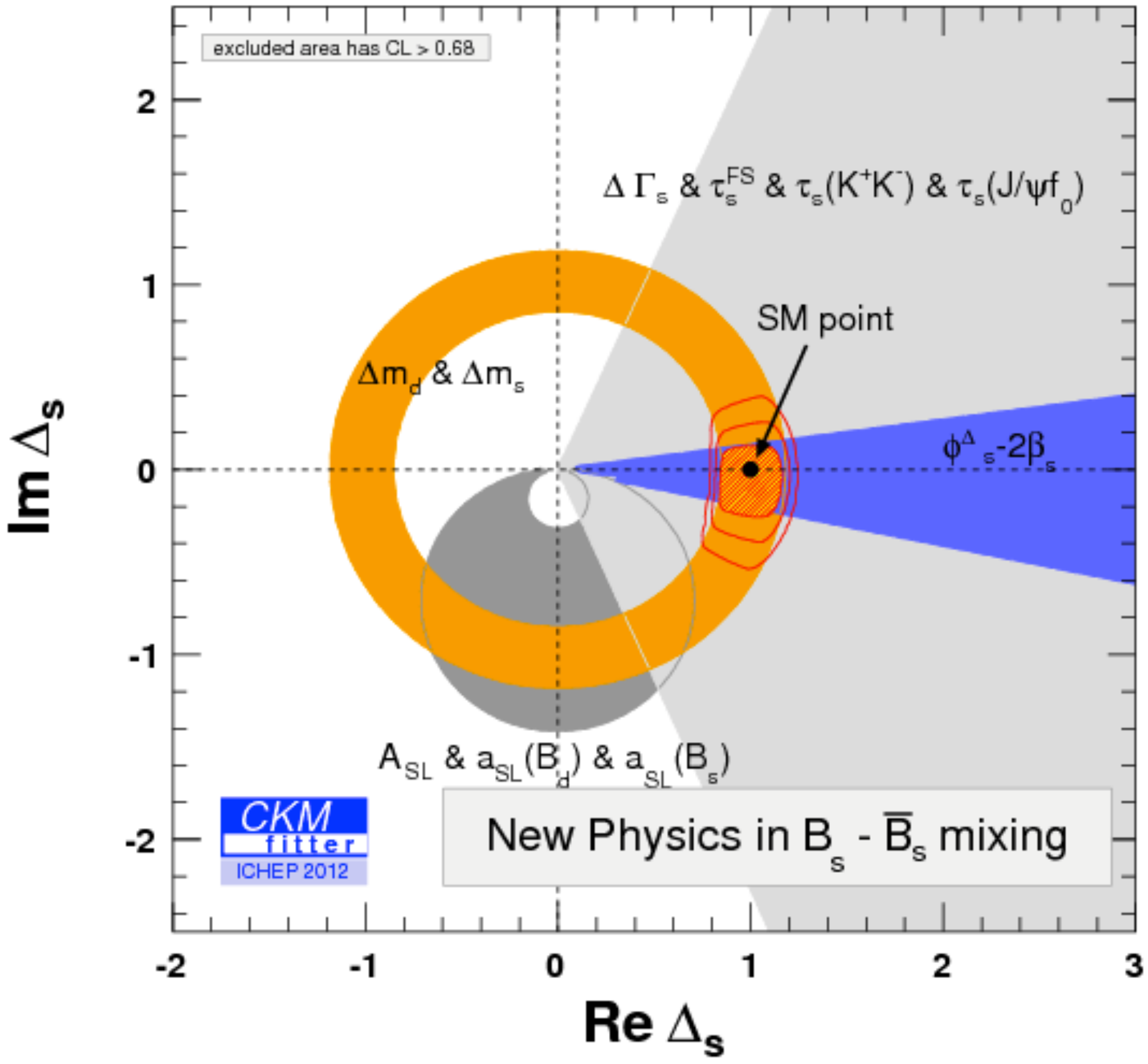}}%
\vspace*{-1ex}
\caption{Present CKM fitter constraints on $\Delta_q\equiv1+h^B_q$ for
  $q=d,s$. From~\cite{Charles:2005ckm}.\label{fig:ckmfitter}}
\end{figure}

The up to date experimental constraints from $B$-meson oscillations from the data fit are
reported in~\cite{Charles:2005ckm} and graphically in Fig.~\ref{fig:ckmfitter} in terms of
$\Delta_q\equiv 1+h^B_q$. While the $B_s$ data agree impressively with the SM, a marginal 1.5\,$\sigma$ CP
deviation still remains in the $B_d$ data. In the following we shall fit the LR $\Delta F=2$ amplitudes within the given $\sigma$-contours and exhibit  the correlated constraints on the relevant mass scales and mixing parameters.

A recent discussion of the SM prediction of $\varepsilon_K$ and the related uncertainties is found
in Ref.~\cite{Buras:2013ooa}.  We will conservatively allow $\mod{h^K_{\varepsilon}}$ to vary within a 20\% symmetric range.

More uncertain is the SM prediction of $\Delta M_{K}$, with equal sharing among short-distance (SD)
and long-distance (LD) theoretical uncertainties as we recap in the following section.

\subsection{Theoretical uncertainties in  $\Delta M_K$}\label{sec:DMuncertainties}

\begin{figure*}[t]
\centerline{%
\includegraphics[width=.97\columnwidth]{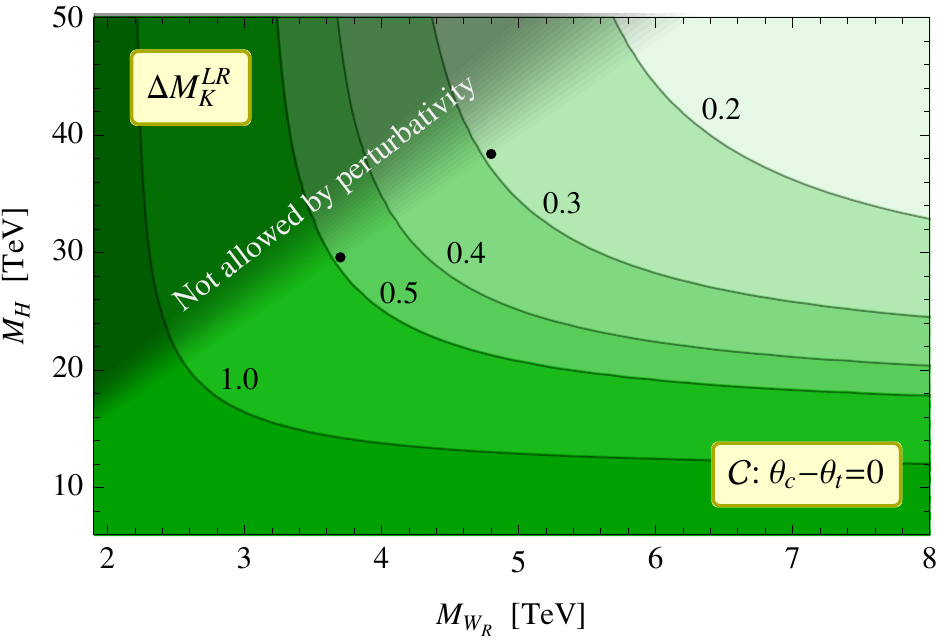}~~~~
\includegraphics[width=.97\columnwidth]{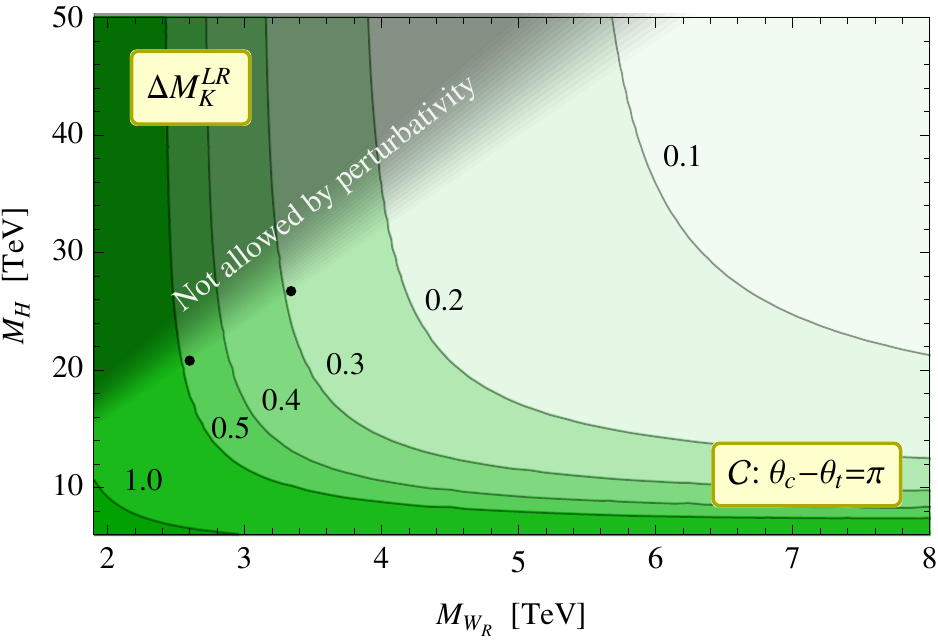}%
}%
\vspace*{-2ex}
\caption{Correlated lower bounds on $M_{W_R}$ and $M_H$ from $\mod{\Delta M_K^{LR}}/\Delta
  M_K^{exp}<1.0,...,0.1$ and for $\theta_c-\theta_t=$ 0 (left) or $\pi$ (right). The latter
  respectively denote constructive and destructive interference between the leading $cc$ and the
  $ct$ amplitudes.\vspace*{0ex}}
\label{fig:K}
\end{figure*}

\noindent
The up to date SD prediction of $\Delta M_K$ within the SM falls just short of the experimental
value and amounts to $(0.9\pm 0.3) \Delta
M_K^{exp}$~\cite{Herrlich:1993yv,Herrlich:1996vf,Brod:2010mj,Brod:2011ty}, where the error is mainly
due to the large uncertainty exhibited by the SM $\eta_{cc}=1.87(76)$ parameter which is now
available at the next-to-next-to-leading order (NNLO)~\cite{Brod:2011ty} in the QCD leading log
resummation. The size of this crucial parameter has increased by 36\% compared to the NLO
calculation, bringing the SD contribution in the ballpark of the experimental value (with some worry
on the convergence of the expansion accounted for in the large error).

On the other hand, it is well known that potentially large LD contributions have to be included as
well. A very recent reassessment of such a LD contributions in the large N expansion is presented in
Ref.~\cite{Buras:2014maa}. By including $1/N$ corrections the authors find $\Delta M_K^{LD}=(0.2\pm
0.1) \Delta M_K^{exp}$.

In comparison, by considering the leading pion exchange and the tree-level $\eta'$ contribution one
obtains $\Delta M_K^{LD}=(0.1\pm 0.2) \Delta M_K^{exp}$~\cite{Buras:2010pza}, where the uncertainty
is driven by the meson loop scale dependence.


A chiral quark model prediction of $\Delta M_K^{LD}$ at the NLO in the chiral expansion was
performed in Ref.~\cite{Antonelli:1996qd,Bertolini:1997ir}.  A quite stable prediction $\Delta
M_K^{LD}\approx -0.1\ \Delta M_K^{exp}$ was found, where the negative value is driven by
non-factorizable gluon condensate corrections to the $\Delta S=1$ chiral coefficients.

Quite recently a full lattice calculation of $\Delta M_K$ on a 2+1 flavor domain wall fermion, has
appeared~\cite{Christ:2012se,Yu:2013qfa} that accounts for $(0.95\pm 0.1) \Delta M_K^{exp}$. Such a
result, taken at face value (the quoted error is statistical), very tightly constrains new physics
contributions. On the other hand it is obtained with kinematics still away from physical and
further developments are called for.

\smallskip

In view of the distribution and the theoretical uncertainties related to the SD and LD components we
may conservatively consider a 50\% range of $\Delta M_K^{exp}$ still available for new physics
contributions, even though the recent SD and LD developments hint to a smaller fraction of the
experimental value.

\subsection{Numerical results}\label{subsec:results}

\noindent
In this section we conveniently use the parameters $h$ in \eqs{hKm}{hBq} in order to apply the
experimental constraints on the new physics contributions and to obtain the corresponding bounds on
the LR scales.  Such bounds are set as a correlated constraint on the $M_{W_R}$--$M_H$ plane, once
the relevant LR parameters are marginalized. We should keep in mind that the heavy Higgs $H$ cannot
be decoupled while keeping its couplings to the LR would-be-Goldstone bosons perturbative. Just by
naive dimensional inspection of the effective coupling one must require $M_H/M_{W_R} < 10 $ (a
better, process dependent, assessment based on the convergence of the perturbative expansion
confirms such an expectation~\cite{Basecq:1985cr}). In the following we choose to remain safely
within the non-perturbative regime and exclude the region of $M_H$ above $8 M_{W_R}$ denoted by a
gray smoothed shading in the plots. We discuss separately the $\C$ and $\P$ scenarios.

\subsubsection{Low scale Left-Right $\C$-conjugation}

\paragraph{$\Delta s = 2$ observables.}
We shall begin our discussion with the observables related to $K^0-\overline{K^0}$ mixing.  The
impact of the vertex and self-energies diagrams in Fig. \ref{fig:Diagrams} is for the CP conserving
observable $\Delta M_K$ accidentally low, ranging from 10 to 20\% of the LR box amplitude. This is
well understood because of the $\log(x_c)$ enhancement in the box loop function $F_A(x_c,x_c)$ (see
App.~\ref{sec:loopfunctions}), not present in the vertex and self-energy amplitudes.

On the other hand, for the $ct$ and $tt$ components one expects the vertex and self-energy
amplitudes to be similar in size to the corresponding LR box amplitude and they play indeed a
crucial role, since they add up coherently to the box and tree amplitudes.  This feature holds
independently of the heavy Higgs mass, since, as already discussed, there are components of the $C$
and $D$ amplitudes that do not depend on the Higgs mass (they are in fact needed for the cancelation
of the gauge dependence of the box diagram~\cite{Basecq:1985cr}).

In Fig.~\ref{fig:K} we present the constraints due to the LR contributions to $\Delta M_K$, whose SM
prediction and related uncertainties were summarized in section~\ref{sec:DMuncertainties}.  The
figures are correlated plots in the $M_{H}\text{--}M_{W_R}$ plane for the two phase configurations
$\theta_c-\theta_t=0$ or $\pi$, which lead to constructive or destructive interference between the
$cc$ and $ct$ contributions.

\pagebreak[3]

The destructive interference between the $cc$ and $ct$ amplitudes is now much more effective when
compared to the results of Ref.~\cite{Maiezza:2010ic}. This is a combined effect of the presence of
the additional vertex and self-energy amplitudes and of the proper evaluation of $ \eta_{K ct,tc}$
in \Table{table:eta}.  The latter were underestimated by a factor of four in~\cite{Maiezza:2010ic}.

As a result, the case of $\theta_c-\theta_t=\pi$ (right plot in Fig.~\ref{fig:K}) leads to the more
favourable case: one infers $M_{W_R}>2.6\,(3.4)$TeV when one allows for a 50\,(30)\% LR contribution
to $\Delta M_K$ (see the discussion in Sect.~\ref{sec:DMuncertainties}).

\medskip

The analysis of indirect CP violation in $K$ oscillations, characterized by $\varepsilon$, leads to
important results.  In the case of $\C$-conjugation the dominant LR contributions to the
$h^K_\varepsilon$ can be written in the form
\be
\label{eq:epsC}
h^K_{\varepsilon}\simeq{\rm Im}\!\left[e^{i(\theta_d-\theta_s)}\big( A_{cc}+\!A_{ct}\cos(\theta_{c}-\theta_{t}+\phi)\big)\right],
\ee
where $\phi=\arg(V_{Ltd})\simeq -22^\circ$. $A_{cc,ct}$ are to an extremely good approximation real
numbers (we suppressed the minor $tt$ contribution for simplicity).  For $M_{W_R}$ in the TeV range
we obtain $A_{ct}/A_{cc}\simeq 0.45$, with $A_{cc} \simeq 90$. Analogously to the $\Delta M_K$
discussion, the phase difference $\theta_c-\theta_t$ determines the constructive or destructive
interference between the $cc$ and $ct$ amplitudes.

\begin{figure}[t]
\centerline{%
\includegraphics[width=.97\columnwidth]{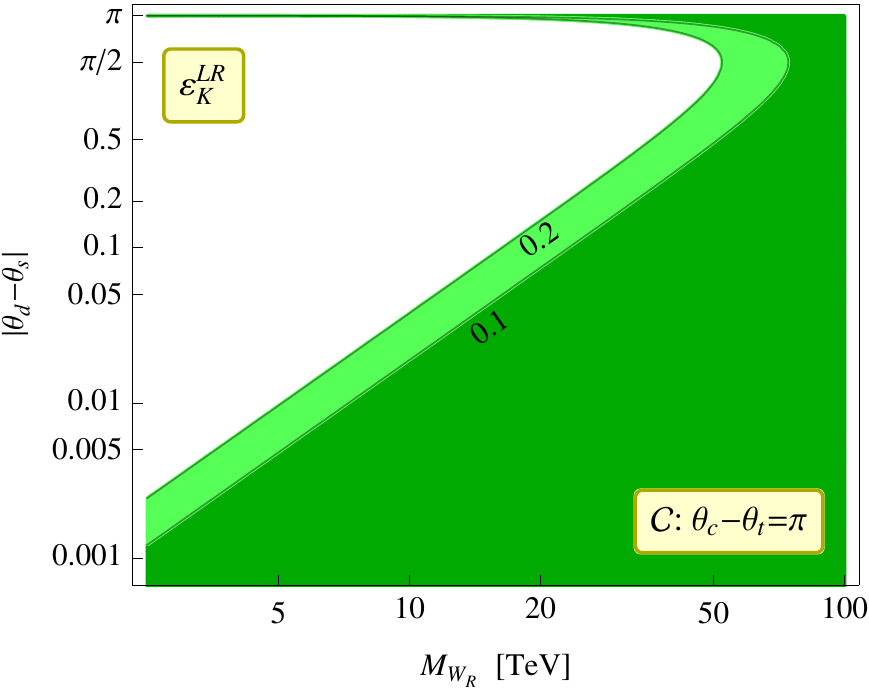}%
}%
\vspace*{-2ex}
\caption{Constraints on the phase $|\theta_d-\theta_s|$ versus $M_{W_R}$, following from the LR
  contributions to $\varepsilon$ in the case of $\C$ as LR symmetry for $\theta_c-\theta_t=\pi$ and
  $h^K_\varepsilon <0.2$ (light shading) and 0.1 (dark shading). For definiteness, we have set
  $M_H=6M_{W_R}$. The plot is periodic for $|\theta_d-\theta_s|\rightarrow |\theta_d-\theta_s|+\pi$.
  \vspace*{-1ex}}
\label{fig:epsC}
\end{figure}

The total amplitudes are sizable and the overall phase $\theta_d-\theta_s$ has to be tuned to reduce
the LR contribution within the allowed limits (we require
$|h^K_\varepsilon|<0.2$~\cite{Buras:2013ooa}).  This means that $\varepsilon$ does not lead to a
bound on $M_{W_R}$ but rather to a constraint on the phase
$\theta_d-\theta_s$~\cite{Zhang:2007da,Maiezza:2010ic}. This is shown in Fig.~\ref{fig:epsC} as a
correlated plot between $\theta_d-\theta_s$ and $M_{W_R}$ for $M_H=6M_{W_R}$. We show the case
relative to $\theta_c-\theta_t=\pi$, the most favorable configuration for low scale LR inferred from
the $\Delta_{M_K}$ discussion, but a very similar result holds for $\theta_c-\theta_t=0$.  From
\eqs{eq:epsP}{eq:epsC} it is clear that plot is periodic in $|\theta_d-\theta_s|$ by $\pi$. The
constraint, evident from the shaded regions in Fig.~\ref{fig:epsC}, is that for $M_{W_R}$ in the TeV
range $|\theta_d-\theta_s|$ has to be very small (see \eq{eq:epsC}), within a few per mil near 0 or
$\pi$.

Regarding $\varepsilon'$, in the minimal LR model one finds, by including the chromomagnetic penguin
contribution~\cite{Bertolini:2012pu} and updated LR matrix elements~\cite{Bertolini:2013noa},
\bea
\varepsilon'_{LR} &\simeq&
|\zeta|\,2.73\big[ \sin(\alpha-\!\theta_u-\!\theta_d)+\sin(\alpha-\!\theta_u-\!\theta_s)\big]\notag\\[1ex]
 && +
|\zeta| \,0.008\big[\sin(\alpha-\!\theta_c-\!\theta_d)+\sin(\alpha-\!\theta_c-\!\theta_s)\big]
\notag\\[1ex]
&& +
\beta\,0.030\sin(\theta_d-\!\theta_s)\,,
\label{epsprimeLR}
\eea
where $\zeta$ is the $W_L$-$W_R$ mixing $\simeq -\beta e^{i\alpha}\tfrac{2x}{1+x^2}$, with $x$ the
modulus of the ratio of the Higgs bi-doublet VEVs, and $\alpha$ their relative phase (see
App.~\ref{appendix:LR}).  The first line in \eq{epsprimeLR} is due to the current-current
$Q_{1,2}^{LR,RL}$ operators, the second to the chromomagnetic LR penguins $Q_{g}^{L,R}$, and the
last line to the current-current $Q_{1,2}^{RR}$ operators (see Ref.~\cite{Bertolini:2012pu} for
notation and details). This expression will be used also below in the case of $\P$.

In the case of $\C$, the conclusion is straightforward: the constraint from $\varepsilon'$ can be
satisfied by either having small enough LR-mixing $\zeta$ (via $x$) or, alternatively, by having
phases $\simeq0$ or $\pi$, which suppress altogether CP violation.  It is indeed a general
fact~\cite{Maiezza:2010ic} that in the case of $\C$ the constraints from CP violation can be
satisfied by the freedom in the CP-phases, thus allowing the LR symmetry at the TeV scale.

\begin{figure*}[t]
\centerline{%
\includegraphics[width=.97\columnwidth]{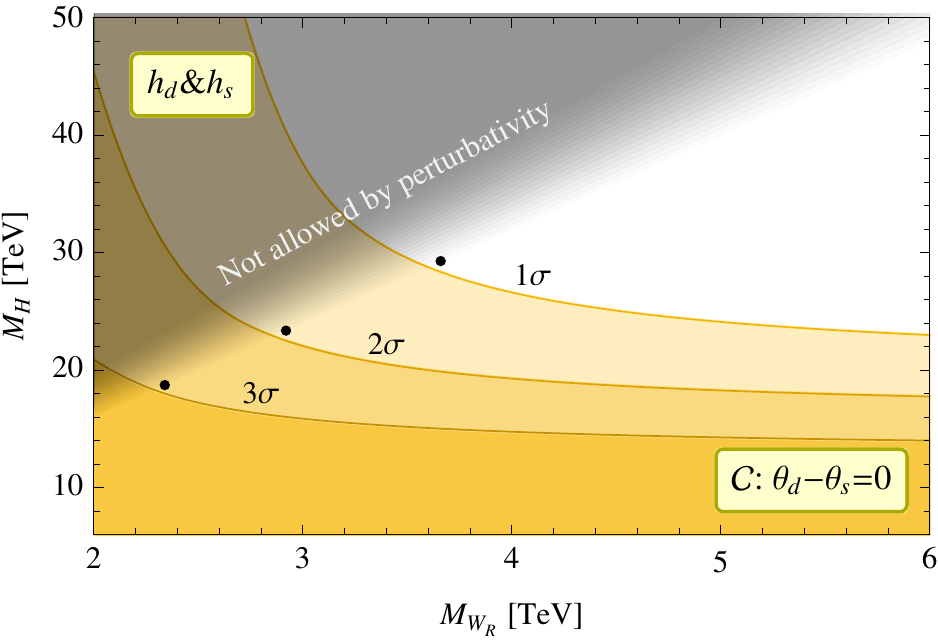}~~~~
\includegraphics[width=.97\columnwidth]{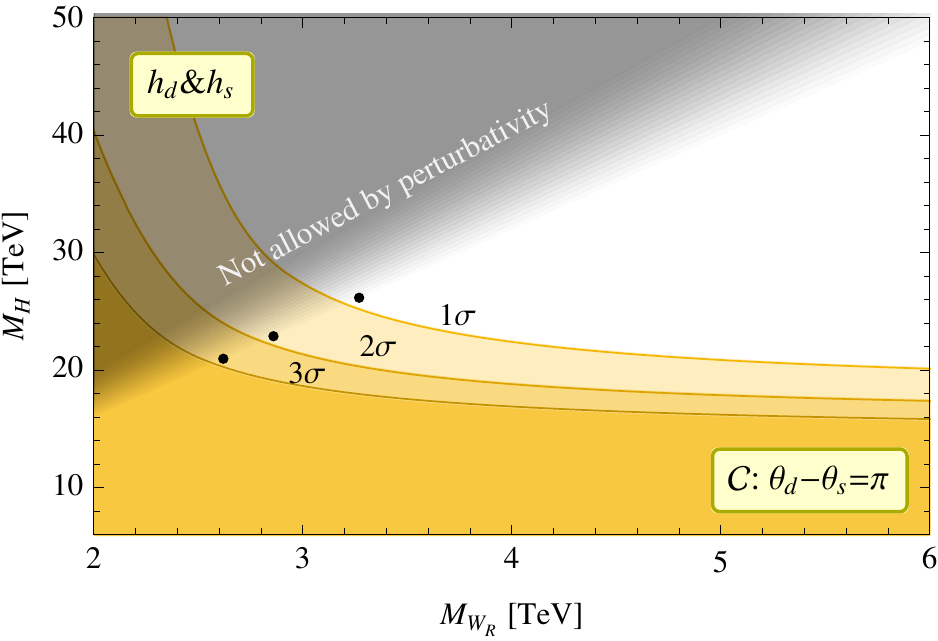}%
}%
\vspace*{-2ex}
\caption{Combined constraints in the $\C$ case on $M_H$ and $M_{W_R}$ from the $B_d$ and $B_s$
  mixings according to the experimental bounds in Fig.~\ref{fig:ckmfitter}, for $\theta_d-\theta_s =
  0$ (left) and $\theta_d-\theta_s=\pi$ (right), as required by $\varepsilon$.\vspace*{-1ex}}
\label{fig:B}%
\end{figure*}

\begin{figure}[t]
\includegraphics[width=.463\columnwidth]{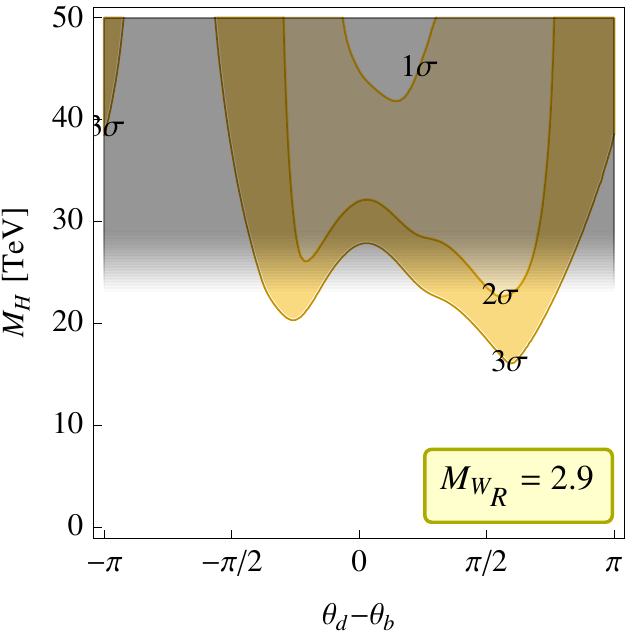}~
\includegraphics[width=.463\columnwidth]{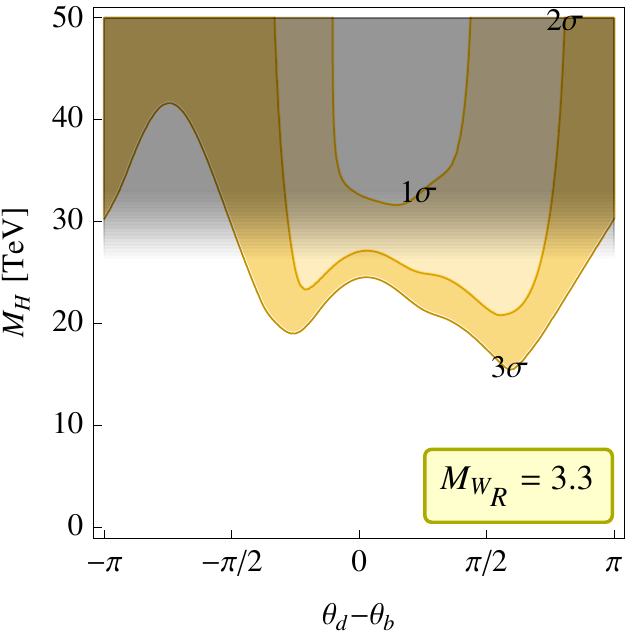}
\includegraphics[width=.463\columnwidth]{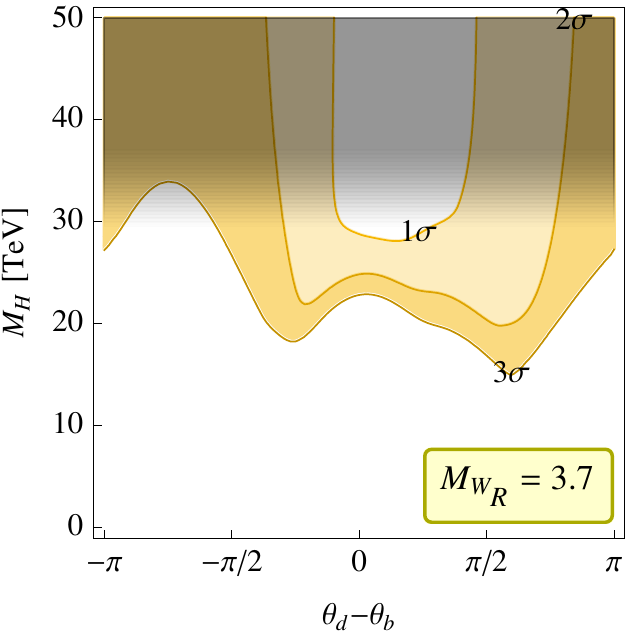}~
\includegraphics[width=.463\columnwidth]{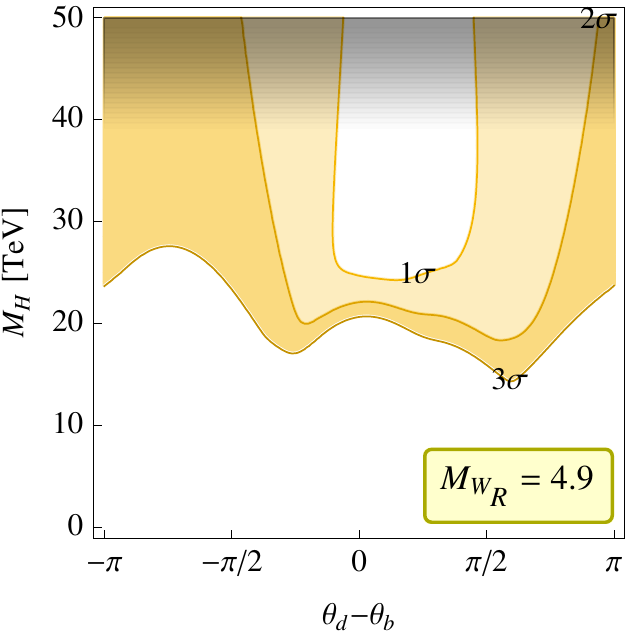}
\caption{Allowed region of $M_H$ and $\theta_d-\theta_b$ (above the contours and below the shading) for  $\theta_d-\theta_s = 0$ and various values of $M_{W_R}$, as obtained from the $B_d$, $B_s$ oscillation data at different confidence levels.}
\label{fig:phase0}
\end{figure}

\begin{figure}[t]
\includegraphics[width=.463\columnwidth]{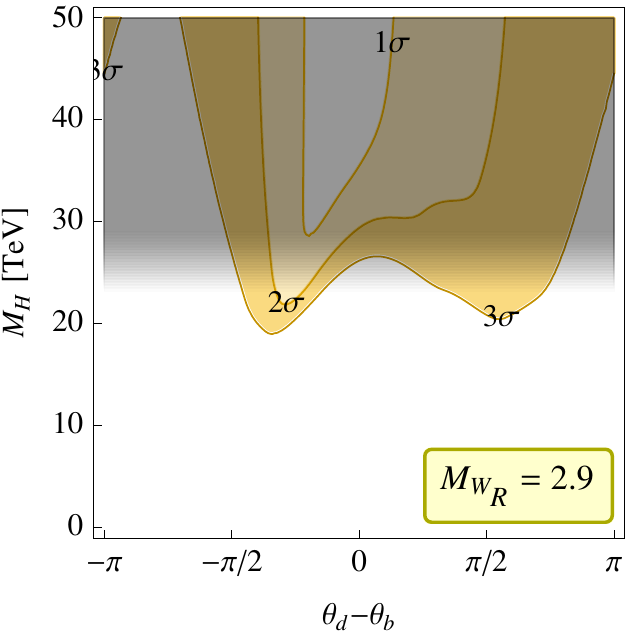}~
\includegraphics[width=.463\columnwidth]{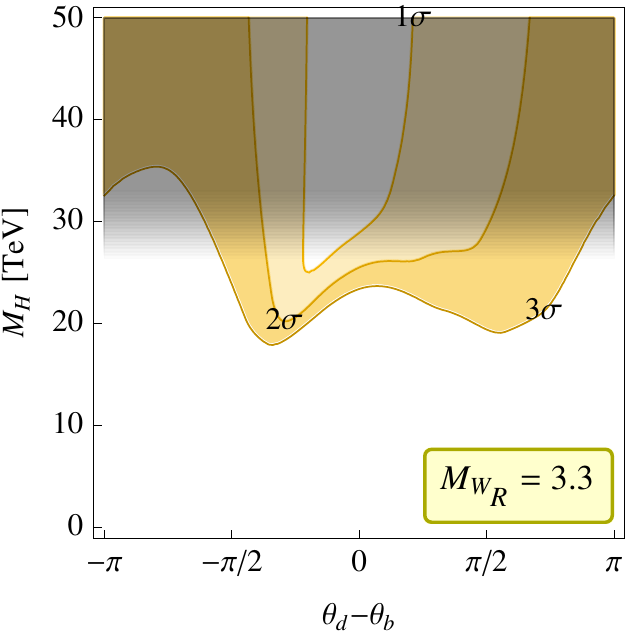}
\includegraphics[width=.463\columnwidth]{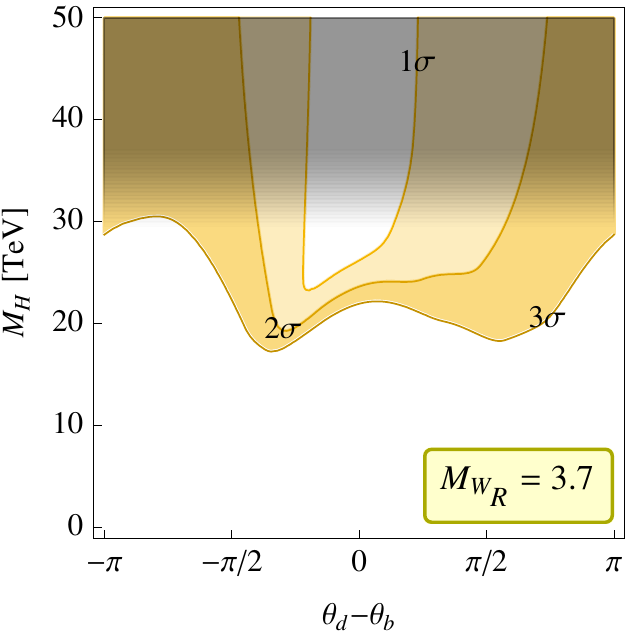}~
\includegraphics[width=.463\columnwidth]{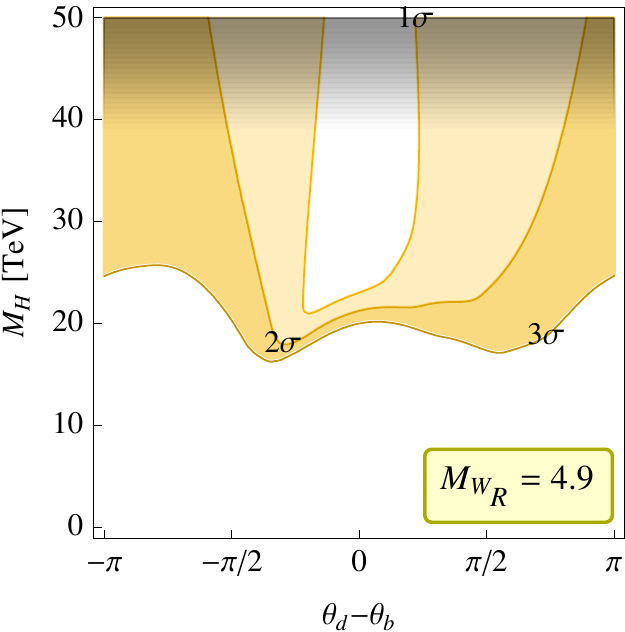}
\caption{Allowed region of $M_H$ and  $\theta_d-\theta_b$ (above the contours and below the shading) for  $\theta_d-\theta_s = \pi$ and various values of $M_{W_R}$, as obtained from the $B_d$, $B_s$ oscillation data  at different confidence levels. }
\label{fig:phasePi}
\end{figure}

\medskip

\paragraph{$\Delta b = 2$ observables.}
The analysis of $B_{d,s}$ mixing is substantially affected by the presence of diagrams $C$ and $D$
of Fig.~\ref{fig:Diagrams}. In fact, the loop amplitudes are dominated by the top quark exchange,
thus the LR box diagram is no longer logarithmically enhanced and the $W_L$-$W_R$ renormalization
diagrams lead to relevant additions, typically of the same order (independently from the heavy Higgs
mass, as we discussed).

It turns out that in spite of the absence of the log and chiral enhancements present in the kaon
case, $B$-mixing is sensitive to LR multiTev scales. In addition to a numerical factor O(10) in the
Wilson coefficient functions, the LR hamiltonian exhibits, when compared to the SM, a factor of four
$\sim m_t^2/m_{W_L}^2$ due to the needed helicity flips on the top quark propagator, and a further
factor of four from the ratio of the QCD factors (compare \Eqs{Ha}{Hd} and the results in
App.~\ref{sec:loopfunctions}).  Finally, the coherent presence of the tree-level FC Higgs
contribution and its one-loop renormalization add, even for a heavy H, another factor 3--4.  All in
all these numerical enhancements make the $\Delta b = 2$ observables sensitive to a $W_R$ mass in
the multi TeV range.  This is a case where a naive analysis based on the relevant effective
operators would be likely misleading.

We analyze the constraints from $B_d$ and $B_s$ mixings by means of the parameters $h_{d,s}$ in
\Eq{hBq}. They are complex and, while their moduli are controlled by $M_H$ and $M_{W_R}$, in the
$\C$-conjugation scenario their phases are related.  By taking into account that
$\theta_d-\theta_s\simeq0,\pi$ from the previous discussion, we can parametrize both $h_{d,s}$ by
the same free combination of phases $\theta_d-\theta_b$, namely:
\be
h_d\sim -{\rm e}^{i(\theta_d-\theta_b-2\phi)}\,,\qquad h_s\sim \mp{\rm e}^{i(\theta_d-\theta_b)}\,,
\ee
where again $2\phi=2\arg{(V_L)_{td}}\simeq -44^\circ$ and the sign $\mp$ follows from
$\theta_d-\theta_s\simeq 0$ or $\pi$ respectively.  The numerical analysis requires to marginalize
over $\theta_d-\theta_b$, by fitting in both $B_d$ and $B_s$ constraints in
Fig.~\ref{fig:ckmfitter}.

Our results are shown in Fig.~\ref{fig:B}, the left and the right plots corresponding to
$\theta_d-\theta_s=0,\pi$ respectively. The latter configuration minimizes the LR scale and we
obtain $M_{W_R}>2.9$--$3.3$\,TeV, at 2 and $1\,\sigma$\,CL respectively.  It is worth mentioning that
presently the experimental data on $B_d$-mixing~\cite{Charles:2005ckm} show a mild $1.5\,\sigma$
discrepancy with the SM prediction and the LR model helps to lighten the tension to below
$1\,\sigma$.

It is remarkable that the bounds on the LR scale from B-physics turn out to be competitive or even
stronger than those obtained from kaon physics. This is due partly to the improvements of the data
and partly to the due inclusion of all relevant contributions, while large LD uncertainties still
affect the SM prediction of $\Delta M_K$.  It is worth mentioning that, since $h_d$ deviates at
present by $1.5\,\sigma$ from 0 (the SM value), the requirement that LR contributions make the theory
consistent with the $1\,\sigma$ experimental region would call for $M_{W_R}<8.0$\,TeV, still in the
limit of large $M_H$.

For either choice $\theta_d -\theta_s=0$, $\pi$ we also scan the allowed ranges of the free model
phase $\theta_d -\theta_b$.  These are shown in \figs{fig:phase0}{fig:phasePi} respectively, in the
$M_H$--$(\theta_d\!  -\!\theta_b)$ plane, for typical values of $M_{W_R}$.  Depending on whether the
LR scale sits onto a minimum or higher, the phase difference is sharply determined or spans a
range. For $M_{W_R}\sim 5\,\TeV$, $\theta_d -\theta_b$ is restricted to vary from 1.5 to 2.5 at the
95\%\,CL.

Overall, the results are summarized in Table~\ref{table:LRboundsC} for two benchmark settings of
$h^K$, $h^B$ and LR phases.  An absolute lower bound of 2.9\,TeV on $M_{W_R}$ emerges at 95\%\,CL.
This confirms the possibility of direct detection of the LR gauge bosons at forthcoming $14\,\TeV$
LHC run, whose sensitivity to $W_R$ is expected to reach the 6\,TeV mass
threshold~\cite{Ferrari:2000sp, Gninenko:2006br}.  Let us remark that the bounds quoted in the Table
are obtained for $M_H\gg M_{W_R}$ (still remaining in the perturbative regime for the Higgs
couplings).

\begin{table}[t]
\vspace*{-1ex}%
$$\begin{array}{cccccc}
\hline
   |h^B_{d,s}|     & \   |h^K_m|\ & \dtheta{c}{t}   &\ \dtheta{d}{s} & \dtheta{d}{b}  & M_{W_R}^{min}\,\text{[TeV]}   \\[0.5ex]
\hline
<\!2\sigma & <\!0.5 & 0   & \simeq 0    & -0.8\div 2.4  & 3.7      \\
        &            &     & \simeq \pi & -1.3 \div 1.8 & 3.7      \\
        &            & \pi & \simeq 0   & \simeq  1.7  & \bf 2.9  \\
        &            &     & \simeq \pi & \simeq  -0.9  & \bf 2.9  \\[1ex]
<1\sigma & <\!0.3 & 0   & \simeq 0   & -0.2 \div 1.5 & 4.9      \\
        &            &     & \simeq \pi & -0.5 \div 0.8 & 4.9      \\
        &            & \pi & \simeq 0   & \simeq   0.5  & \bf 3.7  \\
        &            &     & \simeq \pi & \simeq   -0.7  & \bf 3.3  \\
\hline
\end{array}$$
\vspace*{-2ex}%
\caption{Summary of correlated bounds on the LR scale (in TeV) in the $\C$ case, for two benchmark requirements of $h^K$, $h^B$ and of the relative patterns of LR phases. The limits where the $B$-mixing constraints prevail over $K$-mixing are marked in bold. These represent the most conservative bounds ($M_H \gg M_{W_R}$). The absolute lower bound is $M_{W_R}>2.9\,\TeV$ and there a preferred value of $\dtheta{d}{b}\simeq -1.7$ or 0.9 emerges, depending on $\dtheta{d}{s}$.\label{table:LRboundsC}\vspace*{-1ex}}
\end{table}

\subsubsection{Low scale $\P$-parity}

$\P$ parity in the LR symmetric model requires the Yukawa couplings $Y$ and $\tilde Y$ to be
hermitean (see appendix~\ref{appendix:LR} for notation). The right-handed mixing matrix is given by
$V_R\simeq K_u V_L K_d$ with $K_{u,d}$ diagonal matrices of phases.  On the other hand, any
additional CP violation (i.e. beyond CKM) arises from a non-zero relative phase ($\alpha$) between
the two doublet VEVs, that is the only source of non-hermiticity of the quark mass matrices.  In the
limit of small ratio of the doublet VEVs ($x\equiv v_2/v_1\ll 1$) an analytical solution can be
found~\cite{Zhang:2007da}, and all phases are parametrized in terms of $x\sin\alpha$.  In particular
for $x\ll m_b/m_t$, all LR phases are bounded in a small range about 0 or $\pi$.

\pagebreak[3]

One would be tempted to conclude, as for the $\C$ case, that in this regime $\P$ parity is a viable
setup for low scale LR symmetry.  However, due to the different relation between the left-handed and
right handed mixing matrices, the dependence of observables on the CP phases differs in the $\P$ and
$\C$ schemes. In particular, for $\Delta S=2$ mixing one obtains
\be
\label{eq:epsP}
\sand{K}{^0}{\H_{LR}} \propto e^{i(\theta_d-\theta_s)}\left[ A_{cc}+\!A_{ct}e^{i \phi}\cos(\theta_{c}-\theta_{t}) \right] ,
\ee
while in the case of $B$ mixing the parameters $h_{d,s}$ read
\be
h_d\sim -{\rm e}^{i(\theta_d-\theta_b)}\,,\qquad h_s\sim -{\rm e}^{i(\theta_s-\theta_b)}\,,
\ee
because the CKM phase $\arg(V_{Ltd})$ cancels in the ratio of the LR and SM leading ($tt$)
amplitudes. For $\theta_d-\theta_s \simeq 0$, $\pi$ the complex vectors $h_{d}$ and $h_{s}$ are
approximately aligned.  On the other hand, as far as the direct CP violation parameter
$\varepsilon'$ is concerned, \eq{epsprimeLR} holds in both $\C$ and $\P$ cases since the top
mediated LR amplitudes turn out to be subleading~\cite{Bertolini:2012pu}.

\begin{figure}[t]
\centerline{%
\includegraphics[width=.97\columnwidth]{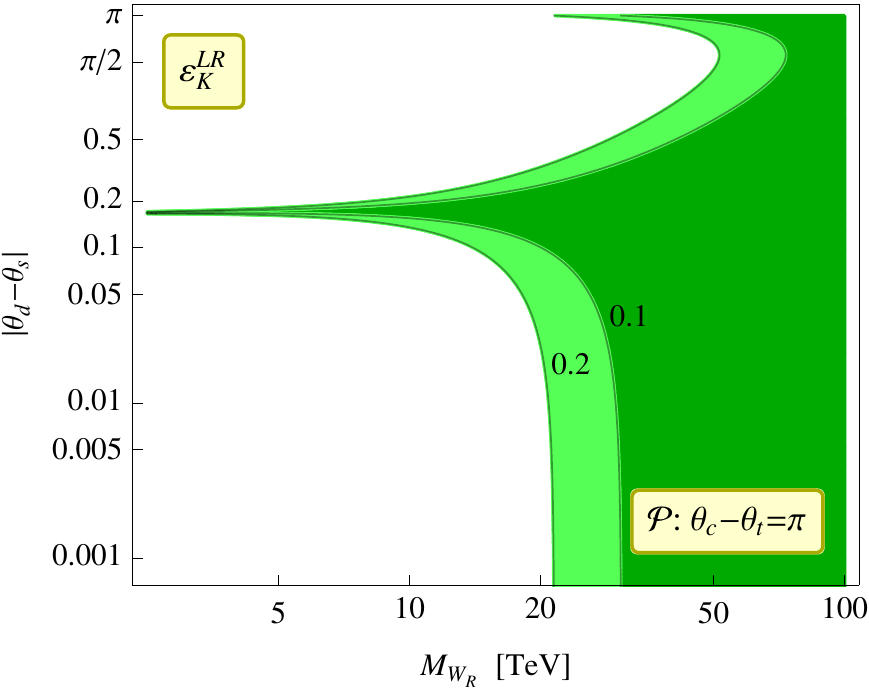}%
}%
\vspace*{-2ex}
\caption{Constraints on the phase $|\theta_d-\theta_s|$ versus $M_{W_R}$, following from the LR
  contributions to $\varepsilon$ in the case of $\P$-parity for $\theta_c-\theta_t=\pi$, and for
  $h^K_\varepsilon <0.2$ (light shading) and 0.1 (dark shading). For definiteness, we have set $M_H=6
  M_{W_R}$. The plot is periodic for $|\theta_d-\theta_s|\rightarrow |\theta_d-\theta_s|+\pi$.
  \vspace*{-1ex}}
\label{fig:epsP}
\end{figure}

\medskip

Let us first discuss the constraint from $\varepsilon$. The interplay in \eq{eq:epsP} of the overall
LR phase $\theta_d-\theta_s$ and the CKM phase in the $ct$ part of the amplitude leads to the
pattern shown in Fig.~\ref{fig:epsP}, to be compared with Fig.~\ref{fig:epsC} in the $\C$ case.  As
it appears, for $M_{W_R}<10\,\TeV$ one is led to the narrow result of $|\theta_d-\theta_s|\simeq
0.17$ (modulo $\pi$).  (An analogous pattern is obtained for $|\theta_c-\theta_t| \simeq 0$.)

This is a fairly large phase, the reason for which being the large ratio $A_{ct}/A_{cc}\simeq 0.45$
in \eq{eq:epsP} combined with the CKM phase $\e^{i\phi}$. The resulting large imaginary part in
$h^K_\varepsilon$ can only be canceled by a fairly large $\theta_d-\theta_s$ phase.

This is a crucial change with respect to the analysis in~\cite{Zhang:2007da,Maiezza:2010ic} where
$A_{ct}/A_{cc}$ resulted much smaller so that the phase $|\theta_d-\theta_s|$ was constrained to be
$\simeq 0$ ,$\pi$ at the percent level.  As already mentioned such a crucially different result has
two distinct and comparable origins: a QCD factor $\eta_{ct}^{LR}$ larger by a factor 4,
underestimated in previous analyses, and the neglected contributions of the self-energy and vertex
renormalization diagrams, which additionally increase the ratio $A_{ct}/A_{cc}$ by approximately a
factor 3.

The consequences of this large phase are important: first, in the regime $x\ll m_b/m_t$ where
analytical expressions for the phases are available~\cite{Zhang:2007da}, it is straightforward to
see using \eq{epsprimeLR} that a strong bound emerges from $\varepsilon'$, which excludes the
scenario of low scale $\P$ LR symmetry.  One finds:
\be
x\ll\frac{m_b}{m_t} \ \ \Rightarrow \ \ \frac{\varepsilon'_{LR}}{\varepsilon'_{exp}}\simeq \left(\frac{10\,\TeV}{M_{W_R}}\right)^2
\ee
which translates into 14\,(10)\,TeV if one tolerates a 50\,(100)\% contribution to $\varepsilon'$.
As a result, one can exclude the regime of hierarchic VEVs $x=v_2/v_1\ll m_b/m_t$ for low scale $\P$
LR symmetry. This has also implications for the analysis of the leptonic sector~\cite{ST}.

\smallskip

\begin{figure}[t]
\includegraphics[width=.97\columnwidth]{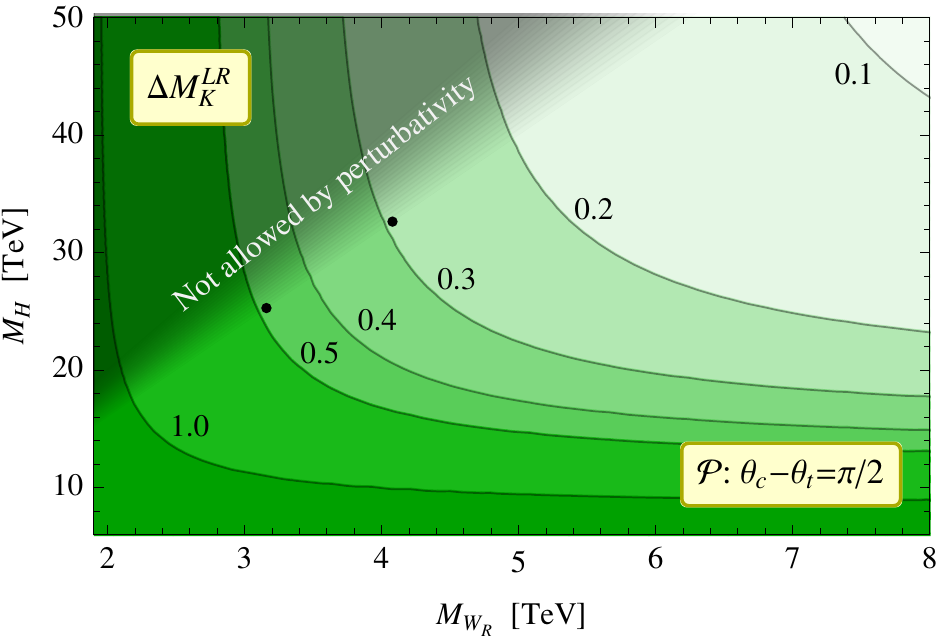}
\vspace*{-2ex}
\caption{Correlated bounds on $M_R$ and $M_{W_R}$ (region above the curves) for $\mod{\Delta
    M_K^{LR}}/\Delta M_K^{exp}<1.0,...,0.1$ and for $\theta_c-\theta_t=\pi/2$ in the case of $\P$
  parity.}
\label{fig:DMKpihalfP}
\end{figure}

\begin{figure}[t]
\includegraphics[width=.97\columnwidth]{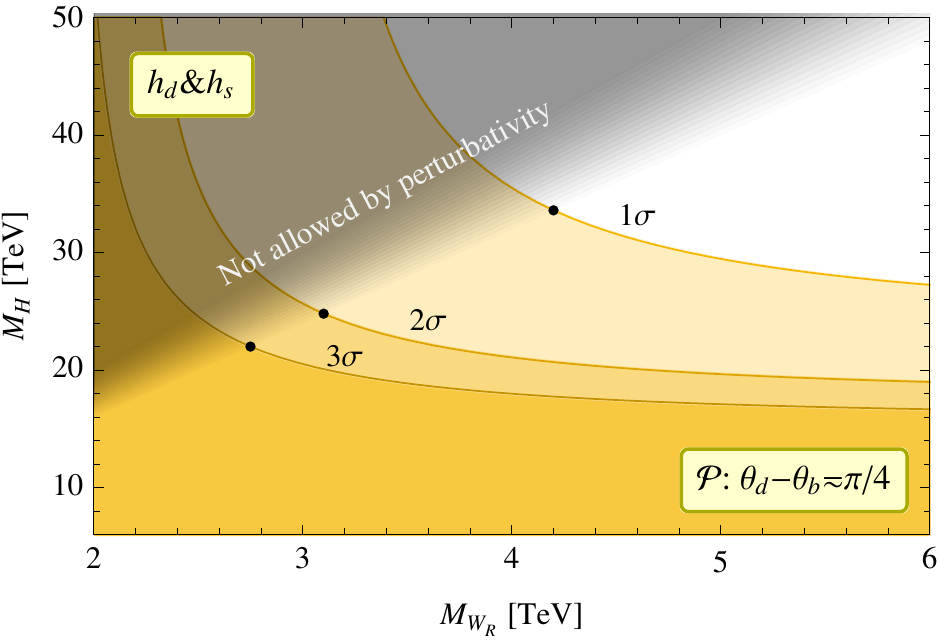}
\vspace*{-2ex}
\caption{Combined constraints on $M_R$ and $M_{W_R}$ from $\varepsilon$,  $\varepsilon'$ $B_d$ and  $B_s$ mixings obtained in the $\P$ parity case from the numerical fit of the Yukawa sector of the model.}
\label{fig:BdBsP}
\end{figure}

On the other hand, when the ratio of the doublet VEVs is larger than a percent, the analytic
solution in~\cite{Zhang:2007da} does not apply, and one expects that for given values of $x$ and
$\alpha$ of order one, the spectrum of the LR phases may exhibit also large values. In order to
address this problem we performed a full numerical analysis of the K and B observables here
discussed. The procedure consists in a $\chi^2$ fit of the known spectrum of charged fermions masses
and mixings, together with the constraints from $\varepsilon$, $\varepsilon'$ and $h_d$, $h_s$ for
the $B$ mesons. The results can be summarized as follows:
\begin{enumerate}

\item We confirm that for small $x< 0.02 \,(0.01) \simeq m_b/m_t$ the model can not accommodate at
  the same time $\varepsilon$ and $\varepsilon'$ (the tension being at $2\,(3)\,\sigma$). This
  confirms our discussion based on the analytic approximation of Ref.~\cite{Zhang:2007da}.

\item The tension is resolved only for larger $x>0.02$. In this case, $x$ becomes also irrelevant
  and good fits can be found regardless of $x$. The solution requires a definite pattern of
  phases: $\dtheta{c}{t}\simeq\pi/2$ (which reduces the imaginary part in \eq{eq:epsP}) together with
  $\dtheta{d}{s}\simeq\pi$ (which is then necessary for $\varepsilon'$, leading to a cancelation
  between the two terms in the first line of \eq{epsprimeLR}).

\item This pattern of phases leads then to a well defined
  bound from $\Delta M_K$ (see \eq{eq:epsP}). This is illustrated in figure~\ref{fig:DMKpihalfP}.

\item $B_d$ mixing data then drive $\theta_d-\theta_b\simeq \pi/4$, see Fig.~\ref{fig:ckmfitter}, where
  the data constraint on New Physics ($h_d$) is weaker.

\item According to this pattern we find $M_{W_R}>3.1$ TeV at 2$\sigma$ C.L. and $M_{W_R}>4.2$ TeV at 1$\sigma$ C.L., as illustrated in \fig{fig:BdBsP}.

\end{enumerate}

\begin{table}[t]
$$\begin{array}{cccccc}
\hline
   |h^B_{d,s}|     & \   |h^K_m|\ &\ |\dtheta{c}{t}|   &\ |\dtheta{d}{s}| & \dtheta{d}{b}  & M_{W_R}^{min}\,\text{[TeV]}   \\[0.5ex]
\hline
<\!2\sigma & <\!0.5 & \simeq \pi/2   & \simeq \pi   & \simeq \pi/4 &  {3.1}\  (\bf 3.2)    \\[1ex]
<\!1\sigma & <\!0.3 &                      &                     &                     &  {\bf 4.2}\  (4.1)  \\
 \hline
\end{array}$$
\caption{Summary of correlated bounds on the LR scale (in TeV) in the $\P$-parity case, for two benchmark requirements on the $h^K$'s and $h^{B}$'s and the favorite pattern of the LR phases. With the given uncertainties the limits  arising from the combined numerical fit of $\varepsilon$, $\varepsilon'$ and $B_{d,s}$ mixings are today competitive with those obtained from $\Delta M_K$ (round brackets).\label{table:LRboundsP}}%
\end{table}

In summary, hierarchic VEVs $x<0.02$ are ruled out for low scale $\P$ LR-symmetry, while for larger
$x$ one can find the allowed region in the $M_H$--$M_{W_R}$
plane, according to \figs{fig:DMKpihalfP}{fig:BdBsP}.  \Table{table:LRboundsP} summarizes the results for the LR scale in the $\P$ case, which we find around $3\,(4)\,\TeV$ for the $2\,(1)\,\sigma$ benchmark settings.

\section{What next?}

In this work we considered the combined constraints on the TeV scale minimal LR model, from $\Delta
F= 2$ observables in $B$ and $K$ physics.  We showed that the meson mixing receives significant
contributions from diagrams that were neglected in past phenomenological analysis, albeit needed for
a gauge invariant result.  The complete calculation together with a more careful assessment of the
relevant QCD renormalization factors leads to two main results: i) the exclusion of the scenario of
hierarchic bidoublet VEVs, $x<0.02$ in the case of $\P$-parity.  ii) the competitive or prevailing
role of B-mixing data in setting the lower bounds on the LR scale. Only a substantial progress in
the calculation of the $K_L$-$K_S$ mass difference, e.g.\ from lattice studies) may bring the $\Delta
S=2$ observable in the forefront.

The results are summarized in Tables \ref{table:LRboundsC} and \ref{table:LRboundsP} for two
benchmark settings of $h^K$, $h^B$ and LR phases.  An absolute lower bound of 2.9\,TeV on $M_{W_R}$
emerges at the 95\%\,CL in the case of $\C$.  This confirms the possibility of direct detection of
the LR gauge bosons at the forthcoming $14\,\TeV$ LHC run, whose sensitivity to $W_R$ is expected to
approach the 6\,TeV mass threshold~\cite{Ferrari:2000sp,Gninenko:2006br}.  Let us remark that the
bounds quoted in the tables are obtained for $M_H\gg M_{W_R}$ (still remaining in the perturbative
regime for the Higgs couplings). In the case of comparable Higgs and gauge boson masses we find a
lower limit always above 20\,TeV.

At present, direct searches at LHC provide bounds on the right-handed W bosons that vary according
to the assumptions on the right-handed neutrinos from 2.0 to 2.9\,TeV~\cite{ATLAS:2012ak,
  CMS:2012wr, Chatrchyan:2014koa}.  It is remarkable that even the most conservative indirect lower
bound from B-meson physics is still competitive with the direct search.

\begin{figure}[t]
\centerline{%
\includegraphics[width=.97\columnwidth]{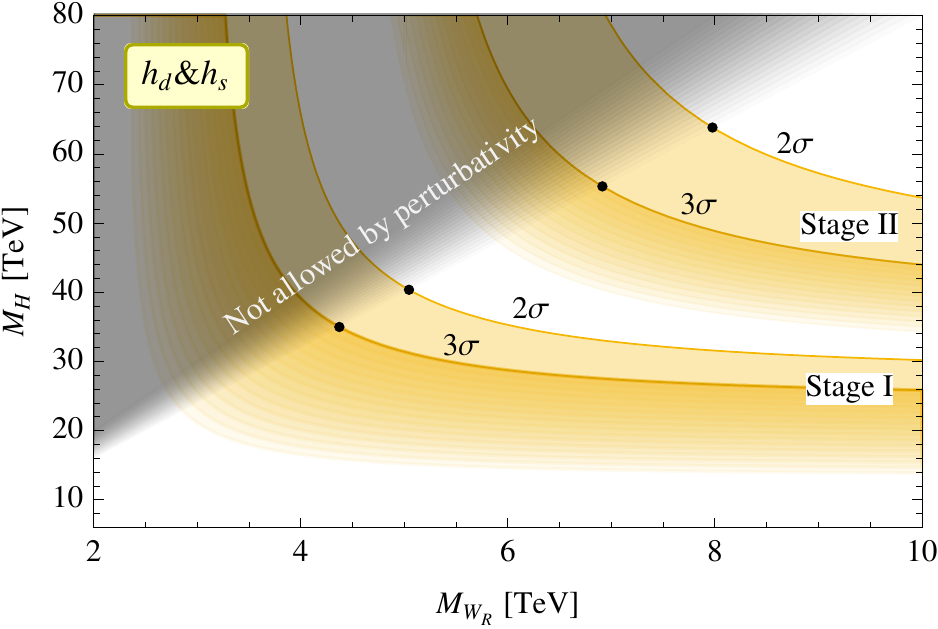}
}%
\vspace*{-2ex}
\caption{Future constraints on $M_R$ and $M_{W_R}$ from the projected combined limits on $h_d$ and
  $h_s$ discussed in Ref.~\cite{Charles:2013aka}. Stage I corresponds to a foreseen 7 fb$^{-1}$ (5
  ab$^{-1}$) data accumulation by LHCb (Belle II) by the end of the decade. Stage II assumes 50
  fb$^{-1}$ (50 ab$^{-1}$) data by the two experiments, achievable by mid 2020's.}
\label{fig:BdBs-Pfit}%
\end{figure}

Sharp improvements in the data are expected from the second LHCb run~\cite{Charles:2013aka}.  The
foreseen data accumulation of LHCb and Belle II in the coming years shall improve on the present
sensitivity by a factor of two within the decade and up to a factor of five by mid 2020s.  The
impact of such an experimental improvement on the sensitivity to the LR scale is depicted in
\fig{fig:BdBs-Pfit}, assuming that the future data on $B_d$ and $B_s$ mixings will be centered on
the SM values.  The shown $\sigma$-contours refer to the foreseen C.L. on the combination of
constraints from $h_d$ and $h_s$. It is noteworthy that the future sensitivity to the LR scale will
reach 7--8\,\TeV, thus exceeding the reach of the direct collider search.

The $B$-physics offers a number of other notable probes of possible new physics, namely rare flavor
changing decays as $B\to\mu^+\mu^-$, $b\to s\gamma$, $b\to s\ \ell^+ \ell^-$, to name a few, and
related CP asymmetries. A comprehensive and updated analysis of the limits on the minimal $\P$ and
$\C$ LR models is currently missing, but a preliminary estimate indicates these processes to be much
less constraining, due to higher backgrounds, less enhancements, or due to the involvement of the
leptonic sector, which still has more freedom in the scales and CP phases. In the arena of indirect
signatures a promising avenue will be the confrontation with electric dipole moments
(EDM). Dedicated efforts are ongoing~\cite{NM,SV} for a reassessment of the limits from nucleon,
atomic and leptonic EDMs.

On the other hand, in the collider arena, in view of the forthcoming high-energy LHC run, an
exhaustive appraisal and exploiting of the various signatures is still timely and compelling in
order to probe the low energy parameter space of $W_R$ and RH neutrinos.

\section*{Acknowledgments}

\noindent

We thank Miha Nemev\v sek, Goran Senjanovic, Vladimir Tello, for useful and stimulating discussions.
S.B. acknowledges partial support by the italian MIUR grant no. 2010YJ2NYW001 and by the EU Marie
Curie ITN UNILHC grant no. PITN-GA-2009-237920.  The work of A.M. was supported in part by the
Spanish Government and ERDF funds from the EU Commission [Grants No. FPA2011-23778,
No. CSD2007-00042 (Consolider Project CPAN)] and by Generalitat Valenciana under Grant
No. PROMETEOII/2013/007.

\appendix

\section{The Left-Right Model}\label{appendix:LR}

\noindent
\paragraph{The gauge lagrangian.}
The minimal LR symmetric extension of the standard electroweak theory is based on the gauge group \cite{Pati:1974yy, Mohapatra:1974hk, Mohapatra:1974gc, Senjanovic:1975rk}
$$
G_{LR}= SU(2)_L \times SU(2)_R \times U(1)_{B-L}\,,
$$
Left and Right quarks and leptons sit in the fundamental representations of $SU(2)_{L,R}$,
%
$Q_{L,R }= \left( u \ d \right)^t_{L,R}$,  
$\ell_{L,R} = \left(\nu \  e \right)^t_{L,R}$,
%
with electric charges $Q = I_{3 L} + I_{3 R} + {B - L \over 2}$, where $I_{3 L,R}$ are the third
generators of $SU(2)_{L,R}$.

The gauge and fermion Lagrangian reads
\begin{eqnarray}\label{lagrangian}
\mathcal{L} &=&i[\overline{\Psi}_{L}\rlap /\!D \Psi_{L}+\overline{\Psi}_{R}\rlap /\! D\Psi_{R}]\\ \nonumber
&-&\frac{1}{4}F_{\mu\nu}F^{\mu\nu}-\frac{1}{4}G_{L\mu\nu}^{i}G_{L}^{i\mu\nu}-\frac{1}{4}G_{R\mu\nu}^{i}G^{i\mu\nu}_{R}
\end{eqnarray}
where $\Psi=(Q\ \ell)$, $\rlap /\!D=\gamma_\mu D^\mu$; $D_{\mu}= \partial_{\mu}-i g W_{\mu
  L}^{i}\frac{\sigma_{L}^i}{2}-i g W_{\mu R}^i \frac{\sigma_{R}^i}{2}-ig' B_{\mu}$ is the $G_{LR}$
covariant derivative ($g_L=g_R=g$).  $F_{\mu\nu} = \partial_{\mu}B_{\nu}-\partial_{\nu}B_{\mu}$ and
$G^{i}_{\mu\nu L,R} = \partial_{\mu}W^{i}_{\nu L,R}-\partial_{\nu}W^{i}_{\mu
  L,R}+g\epsilon^{ijk}W^{j}_{\mu L,R}W^{k}_{\nu L,R}$ are the LR energy-stress tensors.

\paragraph{The Higgs sector.}
The scalar sector contains minimally one right and one left triplet $\Delta_R \in
(1_L,3_R,2)$ and $\Delta_L \in (3_L,1_R,2)$ together with one by-doublet field $\Phi \in (2_L,2_R,0)$~\cite{Senjanovic:1975rk, Senjanovic:1978ev}:
\begin{equation}
\Delta_{} = \left[ \begin{array}{cc} \Delta^+ /\sqrt{2}& \Delta^{++} \\
\Delta^0 & -\Delta^{+}/\sqrt{2} \end{array} \right]_{} , \ \
\Phi = \left[\begin{array}{cc}\phi_1^0&\phi_2^+\\\phi_1^-&\phi_2^0\end{array}\right] .
\end{equation}
The spontaneous symmetry breaking (SSB) of $G_{LR}$ to $SU(3)_C\times U(1)_Q$ is achieved by
\begin{equation}\label{vacuum}
\langle \Delta_L^0\rangle =v_L , \ \
\langle \Delta_R^0\rangle = v_R , \ \
\langle \Phi\rangle = \left[ \begin{array}{cc} v_1 & 0 \\
0 & v_2\, {\rm e}^{i\alpha} \end{array}\right] .
\end{equation}
where $v_L \propto v^2/v_R$ ($v=\sqrt{v_1^2+v_2^2}$). In the broken vacuum we have
\be
\begin{array}{lcl}
M_{W_{L}}^{2} \simeq  g^{2}\frac{v^2}{4}\,, & \quad & M_{Z_{L}}^{2}\simeq \frac{g^{2}}{c_W^2}\frac{v^2}{4} \\[2ex]
M_{W_{R}}^{2} \simeq  g^2 v_R^2 \,, & \quad & M_{Z_{R}}^{2} \simeq  \frac{g^{2}}{c_W^2}\frac{c_W^4|v_{R}|^{2}}{c_W^2-s_W^2}
\end{array}
\ee
where $c_W$, $s_W$ stand for $\cos \theta_W$, $\sin \theta_W$, with $\theta_W$ the Weinberg angle.

The  Yukawa lagrangian reads
\begin{equation}\label{Yukawa}
{\cal L}_Y  = \overline{Q}_L  \Big(Y\, \Phi \,  + \tilde Y \, \tilde\Phi\Big) Q_R + h.c.\,,
\end{equation}
and through the vacuum in \Eq{vacuum} leads to the following quark mass matrices:
\bea
M_u&=&v\, \left(Y \, c  + \tilde Y \, s\, {\rm e}^{-i\alpha} \right)\nonumber\\
M_d&=&v\, \left(Y \, s  \,{\rm e}^{i\alpha}+ \tilde Y \, c \right)\,,
\label{masses}
\eea
where $s=v_2/v$, $c=v_1/v$.  

The diagonal mass matrices $\hat M_q$ are obtained as
\be\label{bidiagonalization}
M_u=U_{uL}\,\hat M_u\,U_{uR}^\dag\,,\qquad M_d=U_{dL}\,\hat M_d\,U_{dR}^\dag\,.
\ee
and the induced flavor mixings in the L, R charged currents are parametrized as
\be
{\mathcal L}_{CC}= \frac{g}{\sqrt{2}}\left[ W_L^\mu \bar{u}_{L\,i} (V_L)_{ij}\g_\mu d_{L\,j}
+  ({L}\to {R}) \right ]+\text{h.c.} \,,
\ee
where $V_{L,R}=U^\dag_{uL,R}U_{dL,R}$ and $i,j$ are flavor indices.

Relevant to our discussion is the presence in the scalar sector of the "heavy" Higgs doublet H. In
terms of the fields $\phi_{1,2}$ it is given by $H=c\phi_2-se^{i \alpha}\phi_1$. From \eq{Yukawa}
one readily shows that the tree level $H^0$-fermion interaction can then be written as:
\bea
\label{FCH}
{\mathcal L}_{H}\simeq  \frac{g}{2M_{W_L}}\bigg[&&
{\bar u}_L\left(V_L\,\hat M_d\,V_R^\dag  \right) u_R \,H^0 \\ \nonumber
&& +{\bar d}_L\left(V_L^\dag\,\hat M_u\,V_R  \right)d_R\, H^{0*}\bigg]+h.c.
\eea
Notice that the $H^0\bar qq$ couplings are proportional to the masses of the opposite isospin quarks.

\paragraph{Discrete LR symmetries.}
The pattern of the left and right mixing matrices is constrained by imposing upon the model a
discrete LR symmetry, which is spontaneously broken together with $SU(2)_R$.  Two realistic
implementations are given by the so called generalized parity $\mathcal{P}$ and conjugation
$\mathcal{C}$ (see Ref.~\cite{Maiezza:2010ic} for a detailed discussion) defined as
\be\label{P&C}
\mathcal{P}: \left\{ \begin{array}{l} Q_L\leftrightarrow Q_R     \\[1ex]  \Phi \to \Phi^\dagger  \end{array}  \right. ,
\qquad
\mathcal{C}: \left\{ \begin{array}{l} Q_L \leftrightarrow (Q_R)^c \\[1ex]  \Phi \to \Phi^T  \end{array}  \right.
\ee
The LR charge-conjugation $\C$ arises naturally in a grand unified embedding as part of the $SO(10)$
algebra.  Imposing $\P$ or $\C$ leads to specific symmetries of the Yukawa couplings and the LR
mixings. In particular one obtains $Y=Y^\dag$ and $Y=Y^T$, respectively. In the same two settings
the mixing matrices are related as in \Eq{eq:VR} respectively by $V_R\simeq K_u V_L K_d$ and
$V_R=K_u V_L^* K_d$, with $K_{u,d}$ diagonal matrices of phases. As recalled in the text, in the
case of $\C$ these phases are free.

In the cases of $\P$ they are all related to a combination of the VEVs ratio $v_2/v_1$ and to
their relative phase $\alpha$. In the limit $v_2/v_1\ll m_b/m_t$ they are numerically of the
order of $m_t/m_b(v_2/v_1)\sin\alpha$~\cite{Zhang:2007da}.

\section{Loop functions}\label{sec:loopfunctions}

For a self-contained discussion we report here
the standard $\Delta F=2$ effective hamiltonian~\cite{Inami:1980fz}
\be\label{HLL}
\mathcal{H}_{LL}^{\Delta F=2}=\frac{G_F^2M_{W_L}^2}{4\pi^2} \sum_{i,j=c,t} \lambda_i^{LL}\lambda_j^{LL} \, \eta^{LL}_{ij} F_{LL}(x_i,x_j) O_{LL}
\ee
where $ \lambda_{i}^{LL}=V^{L*}_{id'}V^L_{id}$, $x_i=m_i^2/M_{W_L}^2$ and
$O_{LL}=\bar d'\g_\mu P_Ld\ \bar d'\g^\mu P_Ld$,
with $\sand{M}{^0}{O_{LL}}=\frac23 f_M^2 m_M$ in the VSA.

The calculation of the QCD renormalization factors $\eta^{LL}_{ij}$ was completed at the NLO in
Ref.~\cite{Herrlich:1996vf} and at the NNLO in Refs.~\cite{Brod:2010mj,Brod:2011ty}. A summary of
updated values is found in Ref.~\cite{Buras:2013ooa}. The loop function can be written as
\be\label{FLL}
F_{LL}(x_i,x_j)=f(x_i,x_j)-f(x_i,0) -f(0,x_j)+f(0,0)
\ee
with
\be
f(x_i,x_j)=\left(1+ \tfrac{x_i x_j}{4} \right) I_2(x_i,x_j,1)
 - 2 x_i x_j  I_1(x_i,x_j,1)
\ee
and
\bea\label{I1}
I_1(x_i,x_j,\beta)&=&\frac{x_i \ln x_i}{(1-x_i)(1-x_i\beta)(x_i-x_j)}+(i\leftrightarrow j) \nn \\
&&- \frac{\beta\ln \beta}{(1-\beta)(1-x_i\beta)(1-x_j\beta)}\ ,
\eea
\bea\label{I2}
I_2(x_i,x_j,\beta)&=&\frac{x_i^2 \ln x_i}{(1-x_i)(1-x_i\beta)(x_i-x_j)}+(i\leftrightarrow j) \nn \\
&&-\frac{\ln \beta}{(1-\beta)(1-x_i\beta)(1-x_j\beta)}\ .
\eea

Finally we list the loop functions appearing in the leading LR Hamiltonians of \eqs{Ha}{Hd}. The loop amplitudes correspond to the results reported in Ref.~\cite{Basecq:1985cr}, with typos amended (and summed over $W_L\leftrightarrow W_R$ exchange). We identify the masses of the scalar and pseudoscalar components of the complex field H, since mass splittings, induced by the electroweak breaking, are negligible compared to the average mass scale.  A convenient subtraction is applied in $\mathcal{H}_ {C,D}$ that identifies $M_H$ with the 1-loop pole mass~\cite{Basecq:1985cr}.

Starting with the LR analogue of the box function in \eq{FLL}, we have in the 't Hooft-Veltman gauge
\be\label{FA}
F_{A}(x_i,x_j,\beta)= \left(1+ \tfrac{x_i x_j\beta}{4}\right) I_1(x_i,x_j,\beta)
 - \tfrac{1+\beta}{4} I_2(x_i,x_j,\beta)\,
\ee
with $\beta=M_{W_L}^2/M_{W_R}^2$, while the self-energy and vertex loop functions in
$\mathcal{H}_{C,D}$ are given by~\cite{Basecq:1985cr}

\begin{widetext}
\bea\label{FC}
F_{C}(M_{W_{L,R}},M_H) &=& \left[10 M_{W_L}^2 M_{W_R}^2+M_{W_L}^4+M_{W_R}^4+M_H ^4\right] \left[I_a(0) - I_a (M^2_H)\right]/M_H ^2  \nn \\
&&+\left[10 M_{W_L}^2 M_{W_R}^2+M_{W_R}^4+M_H ^4\right. -2 M_H ^2 \left.\left(M_{W_L}^2+M_{W_R}^2\right)\right]  I_b(M^2_H)\, ,
\eea

\bea\label{FD}
F_{D}(m_i,m_j,M_{W_{L,R}},M_H) &=& 2 \left(M_{W_L}^2+M_{W_R}^2\right)
\left[I_a(0) - I_a (M^2_H)\right] \nn \\
&&-\left[\frac{M_{W_L} M_{W_R} M_H ^2}{1 - 4 M_{W_L} M_{W_R}/m_i^2} \left[K_a(0,m_i^2)-K_a(M_H^2,m_i^2)\right] + (i\to j)\right]\, ,
\eea
with
\begin{align}\label{IabKa}
&I_a(q^2) = -\frac{i}{\pi^2} \int dk^4 \frac{1}{ (k^2-M_{W_L}^2) [(k+q)^2-M_{W_R}^2]}\\
&I_b(q^2) = -\frac{i}{\pi^2} \int dk^4 \frac{q (k+q)}{q^2 (k^2-M_{W_L}^2) [(k+q)^2-M_{W_R}^2]^2}\\
&K_a(q^2,m_i^2) =
-\frac{i}{\pi^2} \int dk^4 \frac{1}{[(k+\frac{q}{2})^2-m_i^2] (k^2-M_{W_L}^2) [(k+q)^2-M_{W_R}^2]}\, .
\end{align}
\end{widetext}

For convenience we report the results of the integrals in \eq{IabKa} in the relevant limit $m_i^2,M_{W_{L}}^2 \ll M_{W_{R}}^2 \ll M_H^2$:
\begin{align}
\label{lim:Ia0-Ia}
I_a(0) &-\! I_a (M^2_H) \approx \log\frac{M_H^2}{M_{W_R}^2}  -1  
\\[1ex]
\label{lim:Ia}
&I_b (M^2_H) \approx \frac{1}{M_H^2} 
\\[1ex]
\label{lim:Ka0-Ka}
K(0) -\! K(M^2_H) &\approx
\frac{m_i^2 \log\frac{m_i^2}{M_{W_R}^2} -\! M_{W_L}^2 \log\frac{M_{W_L}^2 }{M_{W_R}^2}  }
{M_{W_R}^2\left (m_i^2-M_{W_L}^2\right)}.   
\end{align}
From \eq{lim:Ka0-Ka} and \eq{FD} one readily verifies that the vertex function $F_D$ is fairly insensitive to the quark mass.
The absorptive parts of the $q^2=M_H^2$ subtracted loop integrals are discarded. The dispersive component contributes to the on-shell mass renormalization of the Higgs field. We recall that the chosen subtraction identifies $M_H$ with the one-loop pole mass. 

Since the focus of the authors of Ref.~\cite{Basecq:1985cr} is the study of the cancelation of the gauge dependence their results are not inclusive of the interchange of $W_L$ and $W_R$ in the loops (the cancelation occurs independently in the two sectors). That amounts to an additional factor of two in \eq{FA}  and \eq{FD} (the self-energy amplitude remains invariant). A different overall sign convention is also accounted for.

It is worth noting that in  \eq{FC} and \eq{FD} only the terms that remain proportional to $M_H^2$ are relevant in the heavy $\rm H$ limit. On the other hand, perturbativity of the $H^0 G^+_L G^-_R$ coupling ($G_{L,R}$ are the would-be Goldstone fields) bounds from above the scalar mass  and it conservatively requires $M_H < 8 M_{W_R}$.

\def\arxiv#1[#2]{\href{http:/arxiv.org/abs/#1}{[#2]}}


\begin{thebibliography}{99}

\bibitem{Pati:1974yy}
  J.C.~Pati and A.~Salam,
  ``Lepton Number as the Fourth Color,''
  Phys.\ Rev.\ D {\bf 10}, 275 (1974)
  [Erratum {\bf 11}, 703 (1975)].

\bibitem{Mohapatra:1974hk}
  R.N.~Mohapatra and J.C.~Pati,
  ``Left-Right Gauge Symmetry and an Isoconjugate Model of CP Violation,''
  Phys.\ Rev.\ D {\bf 11}, 566 (1975).

\bibitem{Mohapatra:1974gc}
  R.N.~Mohapatra and J.C.~Pati,
  ``A Natural Left-Right Symmetry,''
  Phys.\ Rev.\ D {\bf 11}, 2558 (1975).

\bibitem{Senjanovic:1975rk}
  G.~Senjanovi\'c and R.N.~Mohapatra,
  ``Exact Left-Right Symmetry and Spontaneous Violation of Parity,''
  Phys.\ Rev.\ D {\bf 12}, 1502 (1975).

\bibitem{Senjanovic:1978ev}
  G.~Senjanovi\'c,
  ``Spontaneous Breakdown of Parity in a Class of Gauge Theories,''
  Nucl.\ Phys.\ B {\bf 153}, 334 (1979).

\bibitem{Minkowski:1977sc}
  P.~Minkowski,
  ``Mu $\to$ E Gamma At A Rate Of One Out Of 1-Billion Muon Decays?,''
  Phys.\ Lett.\ B {\bf 67}, 421 (1977);

\bibitem{Gell-Mann:1980vs}
  M.~Gell-Mann, P.~Ramond and R.~Slansky,
  ``Complex Spinors And Unified Theories,''
  In ``Supergravity'', P.~van Nieuwenhuizen and D.Z.~Freedman (eds.),
  North Holland Publ.\ Co., 1979, p.\ 315;
  Published in Stony Brook Wkshp.\ 1979:0315 (QC178:S8:1979)

\bibitem{Yanagida:1979as}
  T.~Yanagida,
  ``Horizontal Gauge Symmetry And Masses Of Neutrinos,''
  In Proc.\ ``Workshop on the Baryon Number of the Universe
  and Unified Theories'', O.~Sawada and A.~Sugamoto (eds.), Tsukuba, Japan,
  Feb.\ 1979, p.\ 95;

\bibitem{Glashow:1979nm}
  S.L.~Glashow,
  ``The Future Of Elementary Particle Physics,'' HUTP-79-A059
  In Proc.\ Cargese 1979 ``Quarks and Leptons'', p.\ 687;

\bibitem{Mohapatra:1979ia}
  R.N.~Mohapatra and G.~Senjanovi\'c,
  ``Neutrino Mass And Spontaneous Parity Nonconservation,''
  Phys.\ Rev.\ Lett.\  {\bf 44}, 912 (1980).

\bibitem{Feinberg:1978}
  G.~Feinberg, M.~Goldhaber and G.~Steigman,
  ``Multiplicative Baryon Number Conservation and the Oscillation of Hydrogen Into Anti-hydrogen,''
  Phys.\ Rev.\ D {\bf 18}, 1602 (1978).

\bibitem{Mohapatra:1980yp}
  R.N.~Mohapatra and G.~Senjanovic,
  ``Neutrino Masses and Mixings in Gauge Models with Spontaneous Parity Violation,''
  Phys.\ Rev.\ D {\bf 23}, 165 (1981).


 \bibitem{Tello:2010am}
  V.~Tello, M.~Nemevsek, F.~Nesti, G.~Senjanovic and F.~Vissani,
  ``Left-Right Symmetry: from LHC to Neutrinoless Double Beta Decay,''
  Phys.\ Rev.\ Lett.\  {\bf 106}, 151801 (2011)
  \arxiv{1011.3522}[arXiv:1011.3522 [hep-ph]].

\bibitem{Hannestad:2010yi}
  S.~Hannestad, A.~Mirizzi, G.G.~Raffelt and Y.Y.Y.~Wong,
  ``Neutrino and axion hot dark matter bounds after WMAP-7,''
  JCAP {\bf 1008}, 001 (2010)
  \arxiv{1004.0695}[arXiv:1004.0695 [astro-ph.CO]].

\bibitem{Archidiacono:2013cha}
  M.~Archidiacono, S.~Hannestad, A.~Mirizzi, G.~Raffelt and Y.Y.Y.~Wong,
  ``Axion hot dark matter bounds after Planck,''
  JCAP {\bf 1310}, 020 (2013)
  [arXiv:1307.0615 [astro-ph.CO]].

\bibitem{Keung:1983uu}
  W.-Y.~Keung and G.~Senjanovic,
  ``Majorana Neutrinos and the Production of the Right-handed Charged Gauge Boson,''
  Phys.\ Rev.\ Lett.\  {\bf 50}, 1427 (1983).

\bibitem{Nemevsek:2012cd}
  M.~Nemevsek, G.~Senjanovic and Y.~Zhang,
  ``Warm Dark Matter in Low Scale Left-Right Theory,''
  JCAP {\bf 1207}, 006 (2012)
  \arxiv{1205.0844}[arXiv:1205.0844 [hep-ph]].

\bibitem{Nemevsek:2011aa}
  M.~Nemevsek, F.~Nesti, G.~Senjanovic and V.~Tello,
  ``Neutrinoless Double Beta Decay: Low Left-Right Symmetry Scale?,''
  \arxiv{1112.3061}[arXiv:1112.3061 [hep-ph]].

\bibitem{Nemevsek:2012iq}
  M.~Nemevsek, G.~Senjanovic and V.~Tello,
  ``Connecting Dirac and Majorana Neutrino Mass Matrices in the Minimal Left-Right Symmetric Model,''
  Phys.\ Rev.\ Lett.\  {\bf 110}, no. 15, 151802 (2013)
  \arxiv{1211.2837}[arXiv:1211.2837 [hep-ph]].

\bibitem{Zhang:2007da}
  Y.~Zhang, H.~An, X.~Ji and R.N.~Mohapatra,
  ``General CP Violation in Minimal Left-Right Symmetric Model and Constraints on the Right-Handed Scale,''
  Nucl.\ Phys.\ B {\bf 802}, 247 (2008)
  \arxiv{0712.4218}[arXiv:0712.4218 [hep-ph]].

\bibitem{Maiezza:2010ic}
  A.~Maiezza, M.~Nemev\v sek, F.~Nesti and G.~Senjanovi\'c,
  ``Left-Right Symmetry at LHC,''
  Phys.\ Rev.\ D {\bf 82}, 055022 (2010)
   \arxiv{1005.5160}[arXiv:1005.5160 [hep-ph]].

\bibitem{Guadagnoli:2010sd}
  D.~Guadagnoli and R.N.~Mohapatra,
  ``TeV Scale Left Right Symmetry and Flavor Changing Neutral Higgs Effects,''
  Phys.\ Lett.\ B {\bf 694}, 386 (2011)
  \arxiv{1008.1074}[arXiv:1008.1074 [hep-ph]].

\bibitem{Blanke:2011ry}
  M.~Blanke, A.J.~Buras, K.~Gemmler and T.~Heidsieck,
  ``Delta F = 2 observables and B -> Xq gamma decays in the Left-Right Model: Higgs particles striking back,''
  JHEP {\bf 1203} (2012) 024
  \arxiv{1111.5014}[arXiv:1111.5014 [hep-ph]].

\bibitem{Barry:2012ga}
  J.~Barry, L.~Dorame and W.~Rodejohann,
  ``Linear Collider Test of a Neutrinoless Double Beta Decay Mechanism in left-right Symmetric Theories,''
  Eur.\ Phys.\ J.\ C {\bf 72} (2012) 2023
  [arXiv:1203.3365 [hep-ph]].

\bibitem{AguilarSaavedra:2012fu}
  J.A.~Aguilar-Saavedra, F.~Deppisch, O.~Kittel and J.W.F.~Valle,
  ``Flavour in heavy neutrino searches at the LHC,''
  Phys.\ Rev.\ D {\bf 85} (2012) 091301
  [arXiv:1203.5998 [hep-ph]].

\bibitem{Das:2012ii}
  S.P.~Das, F.F.~Deppisch, O.~Kittel and J.W.F.~Valle,
  ``Heavy Neutrinos and Lepton Flavour Violation in Left-Right Symmetric Models at the LHC,''
  Phys.\ Rev.\ D {\bf 86} (2012) 055006
  [arXiv:1206.0256 [hep-ph]].

 \bibitem{Bertolini:2012pu}
  S.~Bertolini, J.O.~Eeg, A.~Maiezza and F.~Nesti,
  ``New physics in $\varepsilon'$ from gluomagnetic contributions and limits on Left-Right symmetry,''
  Phys.\ Rev.\ D {\bf 86}, 095013 (2012)
  \arxiv{1206.0668}[arXiv:1206.0668 [hep-ph]].

\bibitem{Krasnikov:2013ifa}
  N.V.~Krasnikov and V.A.~Matveev,
  ``Search for the right-handed boson $W_{R}$ and heavy neutrino,''
  JETP Lett.\  {\bf 98} (2013) 48
   [Pisma Zh.\ Eksp.\ Teor.\ Fiz.\  {\bf 98} (2013) 53].

\bibitem{Han:2012vk}
  T.~Han, I.~Lewis, R.~Ruiz and Z.-g.~Si,
  ``Lepton Number Violation and $W^\prime$ Chiral Couplings at the LHC,''
  Phys.\ Rev.\ D {\bf 87} (2013) 035011
   [Erratum {\bf 87} (2013) 3,  039906]
  [arXiv:1211.6447 [hep-ph]].

\bibitem{Barry:2013xxa}
  J.~Barry and W.~Rodejohann,
  ``Lepton number and flavour violation in TeV-scale left-right symmetric 
  theories with large left-right mixing,''
  JHEP {\bf 1309} (2013) 153
  \arxiv{1303.6324}[arXiv:1303.6324 [hep-ph]].

\bibitem{Kou:2013gna}
  E.~Kou, C.-D.~Lü and F.-S.~Yu,
  ``Photon Polarization in the $b \to s\gamma$ processes in the Left-Right Symmetric Model,''
  JHEP {\bf 1312} (2013) 102
  [arXiv:1305.3173 [hep-ph]].

\bibitem{Bertolini:2013noa}
  S.~Bertolini, A.~Maiezza and F.~Nesti,
  ``$K$ to $\pi\pi$ hadronic matrix elements of left-right current-current operators,''
  Phys.\ Rev.\ D {\bf 88}, 034014 (2013)
 \arxiv{1305.5739}[arXiv:1305.5739 [hep-ph]].

\bibitem{Dev:2013vba}
  P.S.B.~Dev and R.N.~Mohapatra,
  ``Probing TeV Left-Right Seesaw at Energy and Intensity Frontiers: a Snowmass White Paper,''
  \arxiv{1308.2151}[arXiv:1308.2151 [hep-ph]].

\bibitem{Dev:2013oxa}
  P.S.B.~Dev, C.-H.~Lee and R.N.~Mohapatra,
  ``Natural TeV-Scale Left-Right Seesaw for Neutrinos and Experimental Tests,''
  Phys.\ Rev.\ D {\bf 88}, 093010 (2013)
  \arxiv{1309.0774}[arXiv:1309.0774 [hep-ph]].

\bibitem{Huang:2013kma}
  W.-C.~Huang and J.~Lopez-Pavon,
  ``On neutrinoless double beta decay in the minimal left-right symmetric model,''
  arXiv:1310.0265 [hep-ph].

\bibitem{Roitgrund:2014zka}
  A.~Roitgrund, G.~Eilam and S.~Bar-Shalom,
  ``Implementation of the left-right symmetric model in FeynRules/CalcHep,''
  arXiv:1401.3345 [hep-ph].


\bibitem{Chang:1984hr}
  D.~Chang, J.~Basecq, L.-F.~Li and P.~B.~Pal,
  ``Comment on the $K_L K_S$ Mass Difference in Left-right Model,''
  Phys.\ Rev.\ D {\bf 30}, 1601 (1984).

\bibitem{Basecq:1985cr}
  J.~Basecq, L.-F.~Li and P~B.~Pal,
  ``Gauge Invariant Calculation of the $K_L K_S$ Mass Difference in the Left-right Model,''
  Phys.\ Rev.\ D {\bf 32}, 175 (1985).

\bibitem{Hou:1985ur}
  W.-S.~Hou and A.~Soni,
  ``Gauge Invariance of the $K_L K_S$ Mass Difference in Left-right Symmetric Model,''
  Phys.\ Rev.\ D {\bf 32} (1985) 163.

\bibitem{Ecker:1985vv}
  G.~Ecker and W.~Grimus,
  ``CP violation and left-right symmetry'',
  Nucl.\ Phys.\  B {\bf 258}, 328 (1985).

\bibitem{Chang:1983fu}
  D.~Chang, R.N.~Mohapatra and M.K.~Parida,
  ``Decoupling Parity And SU(2)-R Breaking Scales: A New Approach To Left-Right Symmetric Models,''
  Phys.\ Rev.\ Lett.\  {\bf 52}, 1072 (1984).

\bibitem{Chang:1984uy}
  D.~Chang, R.N.~Mohapatra and M.K.~Parida,
  ``A New Approach To Left-Right Symmetry Breaking In Unified Gauge Theories,''
  Phys.\ Rev.\  D {\bf 30}, 1052 (1984).

\bibitem{Senjanovic:2011zz}
  G.~Senjanovi\'c,
  ``Neutrino mass: From LHC to grand unification,''
  Riv.\ Nuovo Cim.\  {\bf 034} (2011) 1.

\bibitem{Arbelaez:2013nga}
  C.~Arbelaez, M.~Hirsch, M.~Malinsky and J.C.~Romao,
  ``LHC-scale left-right symmetry and unification,''
  \arxiv{1311.3228}[arXiv:1311.3228 [hep-ph]].

\bibitem{Senjanovic:1979cta}
  G.~Senjanovic and P.~Senjanovic,
  ``Suppression of Higgs Strangeness Changing Neutral Currents in a Class of Gauge Theories,''
  Phys.\ Rev.\ D {\bf 21}, 3253 (1980).

\bibitem{Nemevsek:2011hz}
  M.~Nemevsek, F.~Nesti, G.~Senjanovic and Y.~Zhang,
  ``First Limits on Left-Right Symmetry Scale from LHC Data,''
  Phys.\ Rev.\ D {\bf 83} (2011) 115014
  [arXiv:1103.1627 [hep-ph]].

\bibitem{Frere:1991db}
  J.M.~Frere, J.~Galand, A.~Le Yaouanc, L.~Oliver, O.~Pene and J.C.~Raynal,
  ``K0 anti-K0 in the SU(2)-L x SU(2)-R x U(1) model of CP violation,''
  Phys.\ Rev.\ D {\bf 46}, 337 (1992).

\bibitem{Buras:2001ra}
  A.J.~Buras, S.~Jager and J.~Urban,
  ``Master formulae for Delta F=2 NLO QCD factors in the standard model and beyond,''
  Nucl.\ Phys.\ B {\bf 605}, 600 (2001)
  \arxiv{0102316}[hep-ph/0102316].

\bibitem{Vysotsky:1979tu}
  M.I.~Vysotsky,
  ``K0 Anti-k0 Transition In The Standard Su(3) X Su(2) X U(1) Model,''
  Sov.\ J.\ Nucl.\ Phys.\  {\bf 31}, 797 (1980)
  [Yad.\ Fiz.\  {\bf 31}, 1535 (1980)].


\bibitem{Boyle:2012qb}
  P.A.~Boyle {\it et al.}  [RBC and UKQCD Collaborations],
  ``Neutral kaon mixing beyond the standard model with $n_f = 2+1$ chiral fermions,''
  Phys.\ Rev.\ D {\bf 86} (2012) 054028
  \arxiv{1206.5737}[arXiv:1206.5737 [hep-lat]].

\bibitem{Bertone:2012cu}
  V.~Bertone {\it et al.}  [ETM Collaboration],
  ``Kaon Mixing Beyond the SM from N$_{f}$=2 tmQCD and model independent constraints from the UTA,''
  JHEP {\bf 1303}, 089 (2013)
  [Erratum  {\bf 1307}, 143 (2013)]
  \arxiv{1207.1287}[arXiv:1207.1287 [hep-lat]].

\bibitem{Bae:2013tca}
  T.~Bae {\it et al.}  [SWME Collaboration],
  ``Neutral kaon mixing from new physics: matrix elements in $N_f=2+1$ QCD,''
  Phys.\ Rev.\ D {\bf 88} (2013) 071503
  \arxiv{1309.2040}[arXiv:1309.2040 [hep-lat]].

\bibitem{Lytle:2013oqa}
  A.T.~Lytle {\it et al.}  [the RBC-UKQCD Collaboration],
  ``Kaon Mixing Beyond the Standard Model,''
  \arxiv{1311.0322}[arXiv:1311.0322 [hep-lat]].
  
  \bibitem{Aoki:2013ldr}
  S.~Aoki {\it et al.},
  ``Review of lattice results concerning low energy particle physics,''
 \arxiv{1310.8555} [arXiv:1310.8555 [hep-lat]].

\bibitem{Silva:1996ih}
  J.P.~Silva and L.~Wolfenstein,
  ``Detecting new physics from CP-violating phase measurements in B decays,''
  Phys.\ Rev.\  D {\bf 55} (1997) 5331
  \arxiv{9610208}[arXiv:hep-ph/9610208].

\bibitem{Grossman:1997dd}
  Y.~Grossman, Y.~Nir and M.P.~Worah,
  ``A model independent construction of the unitarity triangle,''
  Phys.\ Lett.\  B {\bf 407} (1997) 307
  \arxiv{9704287}[arXiv:hep-ph/9704287].

\bibitem{Deshpande:1996yt}
  N.G.~Deshpande, B.~Dutta and S.~Oh,
  ``SUSY GUTs contributions and model independent extractions of CP phases,''
  Phys.\ Rev.\ Lett.\  {\bf 77} (1996) 4499
  \arxiv{9608231}[arXiv:hep-ph/9608231].

 \bibitem{Lenz:2006hd}
  A.~Lenz and U.~Nierste,
  ``Theoretical update of $B_s - \bar{B}_s$ mixing,''
  JHEP {\bf 0706} (2007) 072.
   \arxiv{0612167}[arXiv:hep-ph/0612167].

\bibitem{Charles:2005ckm}
  CKMfitter Group (J. Charles et al.), 
  Eur. Phys. J. C41, 1-131 (2005) \arxiv{0406184}[hep-ph/0406184], 
  updated results and plots available at: http://ckmfitter.in2p3.fr

\bibitem{Buras:2013ooa}
  A.J.~Buras and J.~Girrbach,
  ``Towards the Identification of New Physics through Quark Flavour Violating Processes,''
  \arxiv{1306.3775}[arXiv:1306.3775 [hep-ph]].

\bibitem{Herrlich:1993yv}
  S.~Herrlich and U.~Nierste,
  ``Enhancement of the K(L) - K(S) mass difference by short distance QCD corrections beyond leading logarithms,''
  Nucl.\ Phys.\ B {\bf 419}, 292 (1994)
  \arxiv{9310311}[hep-ph/9310311].

\bibitem{Herrlich:1996vf}
  S.~Herrlich and U.~Nierste,
  ``The Complete |delta S| = 2 - Hamiltonian in the next-to-leading order,''
  Nucl.\ Phys.\ B {\bf 476}, 27 (1996)
  \arxiv{9604330}[hep-ph/9604330].

\bibitem{Brod:2010mj}
  J.~Brod and M.~Gorbahn,
  ``$\epsilon_K$ at Next-to-Next-to-Leading Order: The Charm-Top-Quark Contribution,''
  Phys.\ Rev.\ D {\bf 82}, 094026 (2010)
  \arxiv{1007.0684}[arXiv:1007.0684 [hep-ph]].

\bibitem{Brod:2011ty}
  J.~Brod and M.~Gorbahn,
  ``Next-to-Next-to-Leading-Order Charm-Quark Contribution to the CP Violation Parameter $\epsilon_K$ and $\Delta M_K$,''
  Phys.\ Rev.\ Lett.\  {\bf 108}, 121801 (2012)
  \arxiv{1108.2036}[arXiv:1108.2036 [hep-ph]].

\bibitem{Buras:2014maa}
  A.J.~Buras, J.-M.~Gerard and W~A.~Bardeen,
  ``Large N Approach to Kaon Decays and Mixing 28 Years Later: Delta I = 1/2 Rule, $\hat B_K$ and $\Delta M_K$,''
  \arxiv{1401.1385}arXiv:1401.1385 [hep-ph].

\bibitem{Buras:2010pza}
  A.J.~Buras, D.~Guadagnoli and G.~Isidori,
  ``On $\epsilon_K$ beyond lowest order in the Operator Product Expansion,''
  Phys.\ Lett.\ B {\bf 688}, 309 (2010)
 \arxiv{1002.3612}[arXiv:1002.3612 [hep-ph]].

\bibitem{Antonelli:1996qd}
  V.~Antonelli, S.~Bertolini, M.~Fabbrichesi and E.I.~Lashin,
  ``The Physics of K0 - anti-K0 mixing: B(K) and Delta M(LS) in the chiral quark model,''
  Nucl.\ Phys.\ B {\bf 493}, 281 (1997)
  \arxiv{9610230}[arxiv:hep-ph/9610230].

 \bibitem{Bertolini:1997ir}
  S.~Bertolini, J.O.~Eeg, M.~Fabbrichesi and E.I.~Lashin,
  ``The Delta I = 1/2 rule and B(K) at O (p**4) in the chiral expansion,''
  Nucl.\ Phys.\ B {\bf 514}, 63 (1998)
  \arxiv{9705244}[hep-ph/9705244].

\bibitem{Christ:2012se}
  N.H.~Christ, T.~Izubuchi, C.T.~Sachrajda, A.~Soni and J.~Yu,
  ``Long distance contribution to the KL-KS mass difference,''
  \arxiv{1212.5931}[arXiv:1212.5931 [hep-lat]].

\bibitem{Yu:2013qfa}
  J.~Yu,
  ``$K_L$-$K_S$ mass difference from Lattice QCD,''
  \arxiv{1312.0306}[arXiv:1312.0306 [hep-lat]].

\bibitem{ST} G. Senjanovi\'c and V. Tello, {\em in preparation}.

\bibitem{Ferrari:2000sp}
  A.~Ferrari {\it et al.},
  ``Sensitivity study for new gauge bosons and right-handed Majorana neutrinos
  in $p p$ collisions at $s$ = 14-TeV,''
  Phys.\ Rev.\  D {\bf 62} (2000) 013001.

\bibitem{Gninenko:2006br}
  S.N.~Gninenko, M.M.~Kirsanov, N.V.~Krasnikov and V.A.~Matveev,
  ``Detection of heavy Majorana neutrinos and right-handed bosons,''
  Phys.\ Atom.\ Nucl.\  {\bf 70} (2007) 441.

\bibitem{ATLAS:2012ak}
  G.~Aad {\it et al.}  [ATLAS Collaboration], ``Search for heavy neutrinos and right-handed $W$
  bosons in events with two leptons and jets in $pp$ collisions at $\sqrt{s}=7$ TeV with the ATLAS
  detector,'' Eur.\ Phys.\ J.\ C {\bf 72}, 2056 (2012) \arxiv{1203.5420}[arXiv:1203.5420 [hep-ex]].

\bibitem{CMS:2012wr}
  CMS Collaboration (CMS Collaboration for the collaboration) ``Search for a heavy neutrino and
  right-handed W of the left-right symmetric model in pp collisions at 8 TeV'' CERN preprint,
  CMS-PAS-EXO-12-017.

\bibitem{Chatrchyan:2014koa}
  S.~Chatrchyan {\it et al.}  [CMS Collaboration],
  ``Search for W' to tb decays in the lepton + jets final state in pp collisions at $\sqrt{s}$ = 8 TeV,''
    \arxiv{1402.2176}[arXiv:1402.2176 [hep-ex]].

\bibitem{Charles:2013aka}
  J.~Charles, S.~Descotes-Genon, Z.~Ligeti, S.~Monteil, M.~Papucci, K.~Trabelsi,
  ``Future sensitivity to new phys\-ics in $B_d$, $B_s$ and $K$ mixings,''
  \arxiv{1309.2293} [arXiv:1309.2293 [hep-ph]].

\bibitem{Inami:1980fz}
  T.~Inami and C.S.~Lim,
  ``Effects of Superheavy Quarks and Leptons in Low-Energy Weak Processes K(L) --> mu anti-mu, K+ --> pi+ neutrino anti-neutrino and K0--anti-K0,''
  Prog.\ Theor.\ Phys.\  {\bf 65}, 297 (1981)
  [Erratum  {\bf 65}, 1772 (1981)].

\bibitem{NM} A. Maiezza and M. Nemev\v sek, {\em work in progress}.

\bibitem{SV} G. Senjanovic and J.C. Vasquez, {\em work in progress}.

\end{thebibliography}
\end{document}